\documentclass[pra,twocolumn,floatfix,nofootinbib,notitlepage,
  superscriptaddress,longbibliography]{revtex4-1}
\usepackage{macros,orcidlink}         %
\date{\today}

\begin{document}

\title{Holstein–Primakoff spin codes for local and collective noise}

\author{Sivaprasad Omanakuttan\,\orcidlink{0000-0002-6229-7087}}
\email[]{sivaprasadto0811@gmail.com}
\affiliation{\addCQuIC}
\affiliation{\addPandAUNM}
\author{Tyler Thurtell\,\orcidlink{0000-0001-8731-6160}}
\affiliation{\addCQuIC}
\affiliation{\addPandAUNM}
\author{Andrew K. Forbes\,\orcidlink{0009-0003-8730-8007}}
\affiliation{\addCQuIC}
\affiliation{\addPandAUNM}
\author{Vikas Buchemmavari \orcidlink{0000-0002-1592-5626}}
\affiliation{\addCQuIC} 
\affiliation{\addPandAUNM}
\author{Ben Q. Baragiola\,\orcidlink{0000-0003-3566-2955}} 
\affiliation{\addCQCCT}
\affiliation{\addYITP} 
\begin{abstract}
{
Quantum error correction is essential for fault-tolerant quantum computation, yet most existing codes rely on local control and stabilizer measurements that are difficult to implement in systems dominated by collective interactions. Inspired by spin-GKP codes~\cite{spin_GKP}, we develop a general framework for \emph{Holstein-Primakoff spin codes}, which maps continuous-variable bosonic codes onto permutation-symmetric spin ensembles via the Holstein-Primakoff approximation. We show that HP codes are robust to both collective and local-spin noise and propose an explicit measurement-free local error recovery procedure to map local noise into correctable collective-spin errors.
}
\end{abstract}

\maketitle

\section{introduction}

Quantum error correction (QEC) is indispensable for scalable quantum information processing, yet most existing frameworks are tailored to architectures with local control, addressability, and measurement of individual qubits. Many physical platforms instead naturally support collective interactions on spin ensembles, where global operations are readily available but local control is limited. Designing error-correcting codes that are intrinsically compatible with such collective dynamics is therefore a central challenge.

Permutation-Invariant (PI) codes~\cite{pollatsek_Rusaki_PIcodes_2004, Ouyang_permutation_invariant_QEC_PRA_2014}, which encode logical qubits in the permutation-symmetric subspace of an ensemble of $N$ spins, offer a promising path forward. Example PI codes include symmetrized versions of the Shor code~\cite{shorcode,Ruskai_PRL_2000_PI_Shor} and the Bacon-Shor code~\cite{BaconShorcode_PRA_2006,Ouyang_permutation_invariant_QEC_PRA_2014}. More recently examples include PI codes that correct absorption-emission type errors~\cite{aydin2024absorption,Jain2024}, have distance scaling of local errors that outperform surface codes~\cite{omanakuttan2023multispin}, feature encoding in different spin irreps~\cite{sharma2024quantum}, and possess transversal $T$ gates~\cite{kubischta2024permutation} or code-switching enabled universality~\cite{Ouyang_Jing_PI_code_switching_PRXQ_2025}. However, these codes are non-additive~\cite{omanakuttan2023multispin, Gross2021} and thus finding natural and experimentally feasible error correction schemes has been a significant challenge.

In the Holstein-Primakoff (HP) regime, where the ensemble is highly polarized and fluctuations are small, the collective spin in the symmetric subspace becomes locally equivalent to a bosonic mode~\cite{holstein1940}. This correspondence was used in Ref.~\cite{spin_GKP} to map continuous-variable Gottesman-Kitev-Preskill (GKP) codes~\cite{GKP_Gottesman_PRA_2001} to PI spin-GKP codes on a large spin ensemble. Spin-GKP codes were shown to have optimal error-correction properties under collective noise, which preserves the symmetric subspace, but their robustness to local-spin noise processes was not known. 

Inspired by that work, we develop a general framework for the subset of PI codes that satisfies the HP approximation, which we call \emph{Holstein-Primakoff spin codes}. We identify and provide a recipe to construct the natural class of HP spin codes: those that are imported via the HP approximation from known bosonic codes. These HP spin codes inherit the error correction properties of the bosonic codes as resilience to collective spin errors, which we show via the Knill-Laflamme conditions. 

Further, we show that PI spin codes satisfying the Knill-Laflamme conditions for collective spin errors automatically have robustness to \emph{local} spin errors as well. This establishes a direct and general connection between collective error correction and robustness to local decoherence. HP spin codes exhibit even more favorable properties: local noise predominantly transfers population between neighboring total-spin irreducible representations while preserving the structure of the encoded codewords. Logical disturbances induced by local errors are suppressed as the number of spins increases. We demonstrate this behavior analytically and numerically for several representative HP code families: spin-GKP, spin-cat, and spin-binomial. Finally, we exploit self-similarity across irreps to construct a measurement-free local error recovery protocol based on a collective SWAP gadget. This protocol coherently maps local spin errors to collective errors without syndrome measurements or feedforward. These results identify HP spin codes as a promising class of PI codes in systems dominated by collective interactions, as they provide a path toward fault-tolerance without the need for local control.

This article is organized as follows. 
\cref{sec:HP_approximation} reviews the HP approximation and the mathematical framework to describe large spin ensembles undergoing local-symmetric noise processes.
\cref{sec:finding_good_spin_codes} lays out the general framework for translating bosonic codes into the symmetric subspace and proves that any code satisfying the Knill–Laflamme conditions for collective errors inherently protects against local spin errors. 
\cref{sec:Local_noise_in_HPA} examines how local depolarizing noise affects states within the HP regime, showing that spin-coherent, spin-cat, and spin-GKP states retain self-similar structure across neighboring irreps, which underlies their robustness to local decoherence. 
\cref{sec:measurement_free_error_correction} introduces a measurement-free recovery protocol based on collective CNOT gates, demonstrating analytically and numerically that local errors can be coherently transformed into correctable collective ones without syndrome measurements.  Finally, \cref{sec:conclusion} summarizes the main results and discusses future directions for scalable, fault-tolerant quantum computation using HP spin codes.

\section{The Holstein-Primakoff approximation}
\label{sec:HP_approximation}

Consider a single quantum spin of total spin $J$ with associated Hilbert space of dimension $2J+1$.
Spin rotations are generated by angular momentum operators $\hat{J}_i$ satisfying the SU(2) commutation relationship
    \begin{equation} 
    \label{eq:angmomcommutation}
        [\hat{J}_i, \hat{J}_j] = i \varepsilon_{ijk} \hat{J}_k ,
    \end{equation}
where $i,j,k \in \{x,y,z\}$, and $\varepsilon_{ijk}$ is the anti-symmetric Levi-Civita tensor. Ladder operators that raise or lower the $\hat J_z$ spin projection are composed of the above by $\hat{J}_+ \coloneqq \hat{J}_x + i \hat{J}_y$ and $\hat{J}_- \coloneqq \hat{J}_x - i \hat{J}_y$. 

The Holstein-Primakoff (HP) transformation is a representation of  this spin in terms of a single bosonic mode. The spin operators are expressed as 
    \begin{subequations} \label{eq:HPtransformation}
    \begin{align}
        \hat{J}_+ &= \sqrt{2J}\sqrt{1 - \frac{\hat{n}}{2J}} \hat{a}
        \\
        \hat{J}_- &= \hat{a}^\dagger \sqrt{2J} \sqrt{1 - \frac{\hat{n}}{2J} } 
        \\
        \hat{J}_z &= J - \hat{n},
    \end{align}
    \end{subequations}
where $\hat{a}$ and $\hat{a}^\dagger$ are the bosonic annihilation and creation operators, and $\hat{n} = \hat{a}^\dagger \hat{a}$ is the number operator~\cite{holstein1940}. The bosonic operators on the right-hand sides of the equations exactly satisfy the SU(2) commutation relations, and the finite dimension of the spin is mapped to a constraint in Fock space: $0 \leq n \leq 2J$.

When $J$ is large, we restrict our attention to a collection of states $\{ \ket{\psi} \}$ localized near the the maximally polarized state along $\hat J_z$ such that
    \begin{equation} \label{eq:HPcondition_weakmain}
            J - \expt{\hat{J}_z} = \expt{\hat n} \ll J,
    \end{equation}
and the operators in the HP transformation, Eqs.~\eqref{eq:HPtransformation}, can be linearized to give
\begin{subequations} \label{eq:HPAconds}
    \begin{align} 
        \hat{J}_+ & \approx \sqrt{2J} \hat{a},
        \\
        \hat{J}_- & \approx \sqrt{2J} \hat{a}^\dagger \\
        \hat{J}_z &= J - \hat{n}.
    \end{align}
    \end{subequations}
This is known as the \emph{Holstein-Primakoff approximation} and is widely used in quantum optics and condensed matter settings, often to study fluctuations around a classical mean~\cite{holstein1940,Hammerer2010,Correggi2015,Vogl2020,Forbes_Collective_States}. In the HP approximation, the $\hat{J}_x$ and $\hat{J}_y$ spin momentum operators map directly to quadratures of the field, $\hat{q}= \frac{1}{\sqrt{2}}(\hat{a} + \hat{a}^\dagger)$ and $\hat{p}= \frac{1}{i\sqrt{2}}(\hat{a} - \hat{a}^\dagger)$, and vice versa
\begin{equation} \label{eq:HPAquadratures}
    \hat{J}_x \approx \sqrt{J} \hat{q}, 
    \quad \quad
    \hat{J}_y \approx \sqrt{J} \hat{p}.  
\end{equation}

In the HP transformation, the spin eigenstates $\hat{J}_z \ket{J,M} = M \ket{J,M}$, are associated with bosonic Fock states via $\ket{J,J-n} = \ket{n}$ with $n$ restricted to $0 \leq n \leq 2J$, and the bosonic vacuum $\ket{0}$ corresponds with the fully polarized spin state $\ket{J,J}$. In the HP approximation, one may restrict to states that are localized near $\ket{J,J}$ such that $\expt{\hat{n}} = \mathcal{O}(1)$.\footnote{Note that any scaling $\expval{\hat n} < \mathcal O(J)$ allows Eqs.~\eqref{eq:HPtransformation} to converge to Eqs.~\eqref{eq:HPAconds} in expectation as $J \rightarrow \infty$. However, we desire a transverse variance to scale as $\mathcal O(J)$, so we restrict our focus to the tighter condition $\expval{\hat n}=\mathcal O(1)$.} In this regime, the transverse fluctuations are small compared to $J$, 
$\expt{\hat{J}_i^2}/J^2 \ll 1$, and exhibit the scaling
    \begin{equation}
        \expt{ \hat J_i^2} = \mathcal O(J) \quad \text{for} \quad i =x,y , 
    \end{equation}
and for the first moments Cauchy-Schwartz gives $|\expt{ \hat J_i }| \le \expt{ \hat J_i^2 }^{1/2} = \mathcal O (J^{1/2}).$

\subsection{Ensembles of small spins}

Our primary concern is not a single large spin but rather an ensemble of $N$ spin-$\frac{1}{2}$ particles, equivalent to $N$ qubits, living in a Hilbert space of dimension $2^N$. A standard basis to express an arbitrary state of the ensemble is the tensor product basis $\ket{\psi} = \sum_{\mathbf{n}} c_{\mathbf{n}} \ket{n_1} \otimes \ket{n_2} \otimes \dots \otimes \ket{n_N}$, where each basis state is an eigenstate of the single spin operator $\hat{j}^{(n)}_z = \frac{1}{2} \hat{\sigma}_z^{(n)}$. We choose to use the quantum information convention that $\ket{0}$ and $\ket{1}$ are the respective $\pm 1$ basis states for each spin. Another useful basis arises from the irreducible representations, or \emph{irreps}, of SU(2). The $2^N$-dimensional Hilbert space factorizes into $N$ spin-$J$ sectors,
\begin{equation}
    \bigotimes_{n=1}^N \mathcal{H}_{j=\frac{1}{2}}^{(n)} =  
    \bigoplus_{J = J_\text{min}}^{J_\text{max}} \Bigg(  \bigoplus_{\lambda = 1}^{d_N^J} \mathcal{H}^{(\lambda)}_{J} \Bigg),
\end{equation}
where $J_\text{max} = N/2$ and $J_\text{min} = 0$ or $\frac{1}{2}$, for $N$ even or odd. Each irrep has $\dim (\mathcal{H}_{J }^{(\lambda)}) = 2J+1$ and
    \begin{equation}
        d_N^{J}=\frac{N! (2J+1)}{(N/2-J)! (N/2+J+1)!}
        \label{eq:number_of_degeneracy}
    \end{equation}
is the number of degenerate multiplicities labeled by $\lambda$. Thus, any pure state of the ensemble can be written as
    \begin{equation} 
    \label{eq:irrepbasis}
        \ket{\psi} = \sum_{J=J_\text{min}}^{J_\text{max}} \sum_{M = -J}^J \sum_{\lambda = 1}^{d_N^J} c_{J,M,\lambda} \ket{J,M,\lambda}
    \end{equation}
where $0\leq J \leq N/2$ represents the total spin and $-J \leq M \leq J$ is the spin projection along the $z$ axis. 

By design, a property of this basis is that collective spin operators
    \begin{equation} \label{eq:collective_operator}
        \hat{J}_i \coloneqq \frac{1}{2}\sum_{n=1}^{N} \hat{\sigma}_{i}^{(n)},
    \end{equation}
where $i\in \{x,y,z\}$, act independently  on each $\SU(2)$ subspace. 
Unitary and dissipative maps generated by $\hat{J}_i$ act only on the $M$ labels and do not couple irreps or degenerate multiplicities. We will use other properties of the spin-irrep basis, \cref{eq:irrepbasis}, in \cref{sec:Local_noise_in_HPA}, when we consider local-symmetric noise.

For an ensemble of spin-$\frac{1}{2}$ particles with permutation symmetry across all spins, one can describe the system using the collective state space~\cite{chase2008,Baragiola2010,PhysRevA.98.063815} which has dimension $\sim N^2$. 
The collective states spanning this space are invariant under permutations and are defined as~\footnote{We adopt the overline notation from Ref.~\cite{chase2008}; elsewhere it differs.}
    \begin{equation} \label{eq:collective_state}
        \overline{\ketbra{J,M}{ J,M' }} \coloneqq \frac1{d_N^J} \sum_{\lambda =1}^{d_N^J} \ketbra{J,M,\lambda}{J,M^\prime,\lambda}.
    \end{equation}
Though they are not a complete basis for the entire Hilbert space as they cannot describe coherences between different $J$-irreps or between different multiplicities, the set of collective states is preserved under noise that is permutationally symmetric, including local noise---we will rely on this property later.
Further, despite the suggestive notation, the left-hand side of Eq.~(\ref{eq:collective_state}) is not a true outer product of pure states. However, when only describing collective features of the state (such as expectation values of observables in the form of Eq.~(\ref{eq:collective_operator})) one can treat the collective states as an outer product \cite{Forbes_Collective_States}, and thus for the rest of this article we will treat them as such when convenient. Further, collective measurements that cannot distinguish individual spins can only collective-state matrix elements~\cite{sharma2024quantum}.  

Inspired by continuous-variable subsystem decompositions~\cite{Pantaleoni2020,Pantaleoni2023Zak,Shaw2024stabilizer}, one may view the spin-projection degree of freedom and the multiplicity degree of freedom as independent subsystems within each fixed $J$ sector. More precisely, for a given value of $J$, the corresponding Hilbert space factorizes as $\mathcal{H}_J \otimes \mathbb{C}^{d_J^N}$ where collective spin operators act nontrivially only on $\mathcal{H}_J$ and as the identity on the multiplicity space. The collective-state matrix elements defined in Eq.~\eqref{eq:collective_state} are therefore maximally mixed over multiplicities but retain coherence within the spin-$J$ sector. Such mixing does not preclude the use of collective states for quantum-information tasks provided that the encoded information resides entirely within the collective spin degrees of freedom~\cite{Lau2017mixedstate,Marshall2019thermalcomputing}.

In our formulation of HP spin codes below, we focus on the \emph{symmetric subspace}, which is the unique (non-degenerate) irrep corresponding to maximum spin $\Jmax = \frac{N}{2}$, with dimension $N+1$. All states in this subspace satisfy permutation symmetry, and the $\hat{J}_z$-eigenstates in this subspace $\ket{\Jmax, M}$ (degeneracy label dropped), also known as \emph{Dicke states}, are in general entangled states in the individual-spin basis.
One may treat the $N$-spin ensemble as a single spin-$J_\text{max}$ object as long as dynamics and measurements are limited to those generated by collective spin operators, Eq.~\eqref{eq:collective_operator}.

\section{Holstein-Primakoff spin codes}
\label{sec:finding_good_spin_codes}

The HP approximation relating spin systems and harmonic oscillators provides a portal through which bosonic codes can be passed to transform them into spin codes, which we call \emph{Holstein-Primakoff spin codes} (HP spin codes). Such a mapping from continuous-variable (CV) systems to qubits can be done in various ways; for example, a family of HP spin codes, \emph{spin-GKP codes} (reviewed below), was proposed based on mapping CV-GKP codes to a large spin~\cite{spin_GKP}. 

In this work, we provide a general framework that covers spin-GKP codes and, more broadly, describes properties of any spin codes that satisfy the HP approximation. We show that a large class of HP spin codes can be imported directly from CV bosonic codes and we provide the details to do so. Additionally, we focus on ensembles of small spins rather than a single large one. This is because small spins are ubiquitous, whereas high-dimensional spins are not. Some atoms offer relatively large hyperfine spins---\emph{e.g.} Cs-133 has a spin-4 hyperfine ground-state manifold---but even these are minuscule compared to macroscopic ensembles of such atoms. A key property is that PI spin codes, which includes all HP spin codes, have built-in resilience to local-spin errors via the Knill-Laflamme conditions investigated in the next section. Additionally, bosonic codes do not map to local Pauli stabilizer codes; rather, they map to HP spin codes on a ensemble where local addressibility, readout, and control are not required.

\subsection{Knill-Laflamme conditions for collective and local spin errors} \label{sec:KLconditions}

We begin with the error-correcting properties of spin codes.  Since control and readout are restricted to collective spin operators, it is essential to understand which classes of errors can be detected and corrected using only such collective resources.  We analyze the Knill-Laflamme (KL) conditions for permutation-symmetric spin codes under both collective and local error models. 
Generally, the KL conditions for a code that can perfectly correct errors $\{\hat E_a\}$ is summarized by the expression
    \begin{equation} \label{eq:KLconditions}
        \bra{\logic \cwlabel} \hat{E}_a^{\dagger} \hat{E}_b \ket{ \logic \cwlabel'} = \delta_{\cwlabel, \cwlabel'} C_{ab}\, ,
    \end{equation}
where $\ket{ \logic \cwlabel}$ are codewords, and $C_{ab}$ are elements of a Hermitian \emph{QEC matrix}.

Consider an error-correcting code with orthogonal code words, $\inprod{\logic \mu}{ \logic \mu'} = \delta_{\mu, \mu'}$, defined for a spin-$J$ designed to correct a full set of collective spin errors, $\mathcal{E} = \{\hat I, \hat J_x, \hat J_y, \hat J_z\}$. The KL conditions are
    \begin{subequations} \label{eq:KL_angular_momentum}
    \begin{align}
      \bra{\logic{\cwlabel}} \hat{J}_i \ket{\logic{\cwlabel}'}& = \delta_{\cwlabel,\cwlabel'} C_{0i},  \label{eq:KL_angular_momentum_first}
      \\
       \bra{\logic{\cwlabel}} \hat{J}_i \hat{J}_j \ket{\logic{\cwlabel}'} & = \delta_{\cwlabel,\cwlabel'} C_{ij} , \label{eq:KL_angular_momentum_second}
    \end{align}
    \end{subequations}
where $i,j=\{x,y,z\}$ and here the code words are assumed orthogonal.
Now consider this spin-$J$ system to be the symmetric subspace of many spin-$\frac{1}{2}$ particles. 
For the set of spin codes in the symmetric subspace that possess inherent symmetry---such as binary octahedral, binary tetrahedral, and binary dihedral---it was shown that satisfying Eqs.~\eqref{eq:KL_angular_momentum} implies the ability to correct local-spin errors~\cite{omanakuttan2023multispin}. 

Here, we extend this analysis for \emph{any} encoding in the symmetric subspace (any PI spin code), all of which exhibit permutation symmetry by construction. 
Inserting Eq.~\eqref{eq:collective_operator} into \cref{eq:KL_angular_momentum} gives the KL condition,
    \begin{equation} \label{eq:single-spinKL}
        \bra{\logic{\cwlabel}} \hat{\sigma}_i^{(n)} \ket{\logic{\cwlabel}'} = \delta_{\cwlabel,\cwlabel'} \frac{2}{N} C_{0i},
    \end{equation}
where $1 \leq n \leq N$ labels an individual spin. Importantly, this condition is truly local to spin $n$ and does not require that the noise is symmetrized over the ensemble.
For the second-order terms, recall that
    \begin{equation}
        \hat{J}_i \hat{J}_j 
        =
        \frac{N}{4} \delta_{i,j} \hat{I} + \frac{1}{4} \Big( \sum_{n,k} i \epsilon_{ijk} \hat{\sigma}^{(n)}_k + \sum_{n \neq n'} \hat{\sigma}^{(n)}_i \hat{\sigma}^{(n')}_j \Big),
    \end{equation}
with the first summation having $N$ terms and the second $N(N-1)$.
Using Eqs.~(\ref{eq:KL_angular_momentum}--\ref{eq:single-spinKL}) gives the following KL condition on two-body correlations,
    \begin{equation} \label{eq:localspin_KL_correlations}
        \bra{\logic{\cwlabel}} \hat{\sigma}_i^{(n)} \hat{\sigma}_j^{(n')} \ket{\logic{\cwlabel}'}  = \delta_{\cwlabel,\cwlabel'} \left(D _{ij}  - \frac{1}{N-1} \delta_{i,j}\right),
    \end{equation}
when $ n \neq n'$, with coefficients related to those in Eqs.~\eqref{eq:KL_angular_momentum} by
    \begin{equation}
        D_{ij} = \frac{4 C_{ij} -
        2i \epsilon_{ijk} C_{0k}}{N(N-1)} .
        \label{eq:KL_simple_label}
    \end{equation}
The one- and two-body KL conditions, Eq.~\eqref{eq:single-spinKL} and Eq.~\eqref{eq:localspin_KL_correlations}, inherit the property from the collective KL conditions that they do not mix the codewords. 
Thus, a spin code in the symmetric subspace that can correct collective spin errors can also correct local one- and two-body spin errors~\cite{kubischta2024permutation}. Additionally, since local Pauli operators $\{\hat I, \hat\sigma_x, \hat\sigma_y, \hat\sigma_z \}$ form a complete basis for single-spin errors, the above local KL conditions indicate that permutation-symmetric spin codes are intrinsically robust to general local decoherence channels, such as local depolarizing noise. As shown in Appendix~\ref{sec:leakage_error}, this also extends to local leakage errors, which are non-Pauli in nature and relevant for a variety of experimental platforms~\cite{Bluvstein_Lukin_2023_QEC_Logical,Grassl_erasure_1997_PRA,mitra_martin_gate_two_photon,Saffman_review_2016_Rydberg,Sahay_Puri_biased_erasure_PRX_2023,PRXQuantum.5.040343}.  

In many settings, a code cannot correct noise exactly. An example is CV-GKP codes, whose codewords at finite energy are not orthogonal and can at best satisfy the KL conditions approximately for a given noise channel.
The Approximate Quantum Error Correction (AQEC) KL conditions depart slightly from the exact conditions~\cite{approximate_KL_condition},
    \begin{equation}
        \bra{ \logic{\cwlabel} } \hat{ E}_a^\dagger \hat E_b \ket{\logic{\cwlabel}'}
        = \delta_{\cwlabel,\cwlabel'} C_{ab} + \Delta^{\cwlabel \cwlabel'}_{ab},
    \end{equation}  
with the KL remainder $\boldsymbol{\Delta}$ contributing to unavoidable logical error after recovery. If $\| \boldsymbol{\Delta} \|$ is small, the entanglement fidelity of the recovered state obeys $\mathcal{F} \geq 1 - \mathcal O(\Delta^2)$, ensuring that the recovery process remains effective~\cite{approximate_KL_condition}. For PI spin codes, the KL conditions become
    \begin{subequations} \label{eq:AQEC_KL_angular_momentum}
    \begin{align}  \label{eq:AQEC_KL_angular_momentum_a}
        \bra{\logic{\cwlabel}} \hat{J}_i \ket{\logic{\cwlabel}'} & 
        = \delta_{\cwlabel, \cwlabel'} C_{0i} + \Delta^{\cwlabel \cwlabel'}_{0i}, 
        \\
        \bra{\logic{\cwlabel}} \hat{J}_i \hat{J}_j \ket{\logic{\cwlabel}'} 
        &= \delta_{\cwlabel, \cwlabel'} C_{ij} + \Delta^{\cwlabel \cwlabel'}_{ij}, 
    \end{align}
    \end{subequations} 
with an additional component that can arise from non-orthogonal codewords $\Delta^{\mu \mu'}_{00} \coloneqq \inprod{\logic \mu}{\logic{\mu}'}$.
Again, via permutation invariance, the KL remainders in Eqs.~\eqref{eq:AQEC_KL_angular_momentum} are linearly inherited by the local spin KL conditions to give
    \begin{subequations} \label{eq:single-spinKLapprox}
    \begin{align} 
        \bra{\logic{\cwlabel}} \hat{\sigma}_i^{(n)} \ket{\logic{\cwlabel}'} 
        &= \delta_{\cwlabel,\cwlabel'} \frac{2}{N} C_{0i} + \frac{2}{N} \Delta^{\cwlabel \cwlabel'}_{0i} 
        \\
        \bra{\logic{\cwlabel}} \hat{\sigma}_i^{(n)} \hat{\sigma}_j^{(n')} \ket{\logic{\cwlabel}'} 
        &= \delta_{\cwlabel,\cwlabel'} D _{ij} + \tilde{D}_{ij} - \frac{\Delta^{\mu \mu'}_{00}}{N-1} \delta_{i,j}  ,
    \end{align}
    \end{subequations} 
where 
    \begin{equation}
        \tilde{D}_{ij} = \frac{4 \Delta^{\mu \mu'}_{ij}  - 2i \epsilon_{ijk} \Delta^{\mu \mu'}_{0k}}{N(N-1)}.
    \end{equation}
Thus, a PI spin code in the symmetric subspace that approximately corrects collective spin errors can also approximately correct local spin errors.

The results of this section establish that PI spin codes with robustness to collective spin noise automatically inherit robustness to local-spin noise as well.  This observation provides the physical intuition behind the results in Sec.~\eqref{sec:Local_noise_in_HPA}, where we see that local decoherence on HP spin codes primarily transfers population between neighboring total-spin manifolds while preserving the structure of the encoded logical states.  We exploit this property in \cref{sec:Local_noise_in_HPA}
in Sec.~\eqref{sec:measurement_free_error_correction} using a recovery scheme that operates via purely collective spin control.

\subsubsection{Mapping KL conditions from bosonic modes to spins in the Holstein-Primakoff approximation } \label{sec:MappingKLconditions}

The HP approximation provides a systematic way to translate single-mode bosonic codes designed for CV systems~\cite{PhysRevA.97.032346} into HP spin codes in the symmetric subspace. In this mapping, CV error sets for bosonic codes are equivalent to error sets of collective spin operators. As shown in Sec.~\ref{sec:KLconditions}, codes that detect collective spin errors also inherit protection against local spin errors. 

Consider a bosonic code that corrects the error set
\begin{align}
\mathcal{E}_{\mathrm{CV}} =
\{\hat{I}, \hat a, \hat a^{\dagger}, \hat a^{\dagger}\hat a\}.
\end{align}
Via the HP approximation, Eqs.~\eqref{eq:HPAconds}, this error set maps to collective spin errors and additionally includes local one-body and two-body spin errors,
    \begin{equation} \label{eq:HPKLM}
        \mathcal{E}_{\mathrm{spin}}
        = \{\hat{I}, \hat J_{+}, \hat J_{-}, \hat J_z\}
         \cup \{ \text{ local errors } \},
    \end{equation}
with the local errors being suppressed as $N$ grows.

This correspondence extends naturally to higher-order error sets. If a bosonic code is designed to correct polynomial errors in $\hat{a}$, $\hat{a}^\dagger$, and $\hat{n}$, then under the HP mapping, these errors translate to higher powers of collective spin operators and their products. In a spin ensemble, these collective operators decompose into sums of multi-body local spin terms, implying that the imported HP spin code inherits protection against correlated multi-spin errors as well. A full characterization of higher-order errors is beyond the scope here, but the extension is conceptually straightforward: increasing the order of correctable bosonic errors enhances the robustness of the corresponding HP spin code against both collective and local noise.

\subsection{Importing bosonic codes as HP spin codes}

Having established a formal link between the KL conditions for bosonic codes and their HP spin-code counterparts, we now present a constructive recipe to import bosonic codes into HP spin codes. A straightforward way to do so is to begin with a convenient basis for the codewords in the CV setting and then find the associated states in the symmetric subspace of the spins. Two natural bases in the CV setting are the Fock basis and the coherent-state basis, each of which maps cleanly into the symmetric subspace of spin systems via the HP approximation, as we show below.

\subsubsection{Fock basis to Dicke basis}

The CV Fock basis $\{\ket{n}\}$ maps trivially to the Dicke basis in the symmetric subspace over many spins $\{ \ket{J_\text{max},M} \}$, via  $\ket{n} \mapsto \ket{\frac{N}{2}, \frac{N}{2}-n}$. A family of HP spin codes using this mapping arises from binomial codes~\cite{PhysRevX.6.031006}, designed to correct errors generated by $\hat a$, $\hat a^{\dagger}$, and the number operator $\hat n = \hat a^{\dagger}\hat a$, whose codewords are described in the Fock basis. Consider as an example the smallest binomial code designed to correct a single loss. The logical code words
    \begin{subequations} \label{eq:binomialcodewords}
    \begin{align}
        \ket{\logic{0}} &= \tfrac{1}{ \sqrt{2}} (\ket{0} + \ket{4} )
        \\
        \ket{\logic{1}} &= \ket{2} 
    \end{align}
    \end{subequations}
are imported to an HP spin-binomial code with codewords
    \begin{subequations}
    \begin{align}
        \ket{\logic{0}} &= \tfrac{1}{\sqrt{2}} \big(\ket{\tfrac{N}{2}, \tfrac{N}{2}} + \ket{\tfrac{N}{2}, \tfrac{N}{2}-4}\big), 
        \\
        \ket{\logic{1}} &= \ket{\tfrac{N}{2}, \tfrac{N}{2}-2}.
    \end{align}
    \end{subequations}
From Eqs.~\eqref{eq:HPKLM}, the imported HP spin code can correct a single collective $\hat{J}_+$ or local $\hat{\sigma}_+$ spin error approximately, which becomes exact in the $N\to\infty$ limit. 
Binomial codes for higher-order errors in the CV setting can be imported to HP spin codes resilient against higher powers of collective spin operators. Other single-mode rotation symmetric codes~\cite{baragiolaRotation2020} can be imported via same procedure cleanly, although for cat codes importing via the mapping in the following subsection may be preferred.

Under the Fock-to-Dicke mapping, it is important to recognize that HP spin codes can differ from direct encodings into single large spins, with the latter being more general and may not respect the HP approximation.
For example, the small binomial code above can be mapped to a spin code defined on \emph{any} spin of size $J \geq 2$. For small $J$, the imported spin code does not satisfy the HP approximation; for HP spin codes, however, we assume the opposite: $J$ is necessarily large.

\subsubsection{Coherent-state basis to spin-coherent-state basis}
Not all bosonic codes are conveniently described in the Fock basis, with cat codes and GKP codes being prominent examples. For these codes, the overcomplete basis of coherent states $\{ \ket{\alpha} \}$ is more suitable. Coherent states are generated from vacuum via $\ket{\alpha} = \hat{D}(\alpha) \ket{0}$, where $\hat{D}(\alpha)$ are Glauber displacement operators, $\hat{D}(\alpha) \coloneqq e^{\alpha \hat{a}^\dagger - \alpha^* \hat{a}}$ with $\alpha = \alpha_R + i \alpha_I$. 
The HP approximation, Eqs.~\eqref{eq:HPAconds}, associates a spin operator with each Glauber displacement
    \begin{equation} \label{eq:disptorot}
        \hat{D}(\alpha) \approx
         \exp \left[ i \sqrt{\frac{2}{J}} \left(\alpha_I \hat{J}_x - \alpha_R \hat{J}_y\right)\right].
    \end{equation}
The spin operator is a rotation restricted to the $xy$-plane, also called an SU(2) displacement~\cite{akhtar2021subplanck}, defined as
    \begin{equation} \label{eq:SU2displacement}
        \hat{R}(\SCSangle)
        \coloneqq e^{ i \theta \mathbf{n}_\perp \cdot \hat{\mathbf{J}} } = e^{ i\theta (\sin \phi \hat{J}_x - \cos \phi \hat{J}_y  ) },
    \end{equation}
where the axis of rotation is $\mathbf{n}_\perp = (\sin \phi, -\cos \phi, 0) $, and $\SCSangle = \{\theta, \phi\}$ encodes the rotation angles.  For the HP mapping in Eq.~\eqref{eq:disptorot}, the rotation angles in symmetric subspace are given by\footnote{Disentangling the right-hand side of Eq.~\eqref{eq:SU2displacement} provides an alternate Euler-angle description of SU(2) displacements, $\hat{R}(\gamma) = e^{\gamma \hat{J}_-} e^{-\ln (1+ |\gamma|^2) \hat{J}_z} e^{-\gamma^* \hat{J}_+}$, where $\gamma \coloneqq e^{i\phi} \tan \frac{\theta}{2}$. This description can be valuable due to the composition rule $\hat{R}(\gamma_1) \hat{R}(\gamma_2) = \hat{R}(\gamma_3)e^{i \varphi \hat{J}_3}$, where $\gamma_3 = \frac{\gamma_1 + \gamma_2}{1 - \gamma_1^* \gamma_2} $ and $ \varphi  = 2 \arg (1 - \gamma_1^* \gamma_2)$~\cite{akhtar2021subplanck,radcliffe1971properties,Perelmov1972coherentlie}.} 
    \begin{subequations} \label{eq:SCScoeffs}
    \begin{align}
        \phi &= \arg\alpha = \tan^{-1}(\alpha_I/\alpha_R) \\
        \theta &= \tfrac{2}{\sqrt{N}}|\alpha|.
    \end{align}
    \end{subequations}

The relation in Eq.~\eqref{eq:disptorot} provides the link between coherent states and small-angle \emph{spin coherent states} (SCSs). 
Generally, for $N$ spins, a SCS in the symmetric subspace is obtained by applying Eq.~\eqref{eq:SU2displacement} to the maximal $\hat J_z$ projection state,
    \begin{equation} \label{eq:SCS_spinJ}
        \ket{\SCSangle}
         \coloneqq \hat{R}( \SCSangle ) \ket{\tfrac{N}{2},\tfrac{N}{2}}.
    \end{equation}   
Noting that $\hat{R}(\theta,\phi) = \bigotimes_{n=1}^N e^{ i \frac{\theta}{2} ( \sin \phi \hat{\sigma}^{(n)}_x - \cos \phi \hat{\sigma}^{(n)}_y)}$ and that $\ket{\tfrac{N}{2},\tfrac{N}{2}} = \ket{\SCSangle = 0,0} = \ket{0}^{\otimes N}$, a SCS can also be written as the tensor-product state over the spins,
    \begin{equation} \label{eq:qubitSCS}
        \ket{\SCSangle} 
        = \big( \cos \tfrac{\theta}{2}\ket{0} + e^{i\phi}\sin \tfrac{\theta}{2} \ket{1} \big)^{\otimes N}.
    \end{equation}
SCSs are not in general orthogonal~\cite{Arecchi1972},
    \begin{equation} \label{eq:SCSorthogonality}
    \inprod{\SCSangle}{\SCSangle'}
      = 
     [f_{+}(\SCSangle, \SCSangle')]^N
    \end{equation}
with single-qubit overlap function
    \begin{equation} \label{eq:ffunction}
        f_{\pm}(\SCSangle, \SCSangle')
     \coloneqq 
     \cos \tfrac{\theta}{2}
     \cos \tfrac{\theta'}{2} \pm e^{ i (\phi' - \phi)}\sin \tfrac{\theta}{2}
     \sin \tfrac{\theta'}{2} .
    \end{equation}

The angular separation $\Delta \SCSangle \in [0,\pi]$ between two SCSs is the geodesic angle between the Bloch vectors associated with their qubit states from Eq.~\eqref{eq:qubitSCS}, $\mathbf{n}(\SCSangle)$ and $\mathbf{n}(\SCSangle')$, and it satisfies $\cos(\Delta\SCSangle) = \mathbf{n}(\SCSangle)\cdot \mathbf{n}(\SCSangle').$ In terms of the single-spin overlap function, it can be written as
    \begin{align} \label{eq:angularseparation}
        \Delta \SCSangle 
        &= 
        2 \cos^{-1} |f_+(\SCSangle,\SCSangle')| .
    \end{align}
Note that the magnitude of the overlap in Eq.~\eqref{eq:SCSorthogonality} can be written in terms of the angular separation, $| \inprod{\SCSangle}{\SCSangle'}| = (\cos \frac{\Delta \SCSangle}{2} )^N $, which for $\Delta\SCSangle \ll 1$ may be approximated as  $| \inprod{ \SCSangle}{ \SCSangle'} | \simeq \exp \big(- \tfrac{1}{8} N \Delta\SCSangle^{2} \big) $. This reveals a key property: for any fixed nonzero $\Delta \SCSangle$, two SCSs become nearly orthogonal for sufficiently large $N$.
    
The formal link between CV coherent states and SCSs provides the tool to import codes described in this basis. A bosonic codeword described by $\ket{\logic{\mu}} = \sum_k \beta_k \ket{\alpha_k}$ can be directly imported as $\ket{\logic{\mu}} = \sum_k \beta_k \ket{\SCSangle_k}$ with SCS parameters given by Eqs.~\eqref{eq:SCScoeffs}. A simple example is a codeword for a two-lobe cat code $\ket{\logic{\mu}} = \frac{1}{\mathcal{N}} (\ket{\alpha} + \ket{-\alpha})$.
Critical to note is that if the total spin, $J_\text{max} = \frac{N}{2}$, is not large enough, the HP approximation will not be satisfied, and the approximate orthogonality of the two SCSs in superposition will not match that of the CV coherent states. 
It is important to recognize that imported HP spin-cat codewords are \emph{not} GHZ states, which arise in quantum computing \cite{Omanakuttan_spincats_PRXQ_2024,kruckenhauser2024dark,gupta2024universaltransversalgates,yuan2025truncating} and metrology \cite{PhysRevA.100.032318,Huang_2015}. The canonical GHZ state for $N$ spins,
    \begin{subequations} \label{eq:GHZ}
    \begin{align} 
        \ket{\psi_\text{GHZ}}
        &\coloneqq 
        \tfrac{1}{\sqrt{2}} \big(\ket{\SCSangle = 0,0} + \ket{ \SCSangle = \pi,0} \big) \\
        &=
        \tfrac{1}{\sqrt{2}} \big(\ket{0}^{\otimes N}+\ket{1}^{\otimes N} \big)   ,
    \end{align}
    \end{subequations}
has $\expt{\hat{J}_z} = 0$ (as do GHZ states defined with respect to any axis) and, importantly, is not localized near the fully polarized state, so the HP approximation is not valid.

More generally, CV coherent states form a basis, and one is not restricted to discrete superpositions when importing bosonic codes.\footnote{However, for finite $N$, any set of $2J_\text{max}+1 = N+1$ distinct SCSs span the symmetric subspace and form a basis there~\cite{chryssomalakos2018geometry}.} An arbitrary bosonic state may be represented through its Husimi $Q$ function, $Q_{\mathrm{CV}}(\alpha) := \frac{1}{\pi} \bra{ \alpha } \hat{\rho} \ket{ \alpha }$. Under the HP approximation, the $Q$ function admits an analogue in the symmetric spin subspace via the spin $Q$ function, $  Q_{\mathrm{spin}}(\SCSangle) := \bra{\SCSangle} \hat{\rho}  \ket{\SCSangle}.$ For large $N$, these phase-space descriptions are approximately equivalent under the identification of CV phase space with the portion of spin phase space that is locally flat around $\theta = 0$. Consequently, bosonic states---whether expressed as discrete or continuous superpositions---map to spin states whose $Q_\text{spin}$ is concentrated near the north pole of the Bloch sphere. This perspective clarifies the conditions under which the HP approximation remains valid: $Q_\text{spin}$ must be localized to a region that shrinks as $N^{-1/2}$, ensuring that higher-order curvature effects of the Bloch sphere remain negligible.  The Majorana stellar representation of symmetric spin states~\cite{bacry1974orbits, bacry1978minuncertaintystates, karol2021rotosensor, Romero2024constellations} provides a complementary geometric picture: states compatible with the HP approximation have constellations clustered near a single point on the sphere.  Although we do not pursue further analysis, these representations provide useful geometric intuition for understanding the HP approximation and the regime of validity of imported HP spin codes.

\begin{figure*}[t]
    \centering
    \includegraphics[width = 0.96\linewidth]{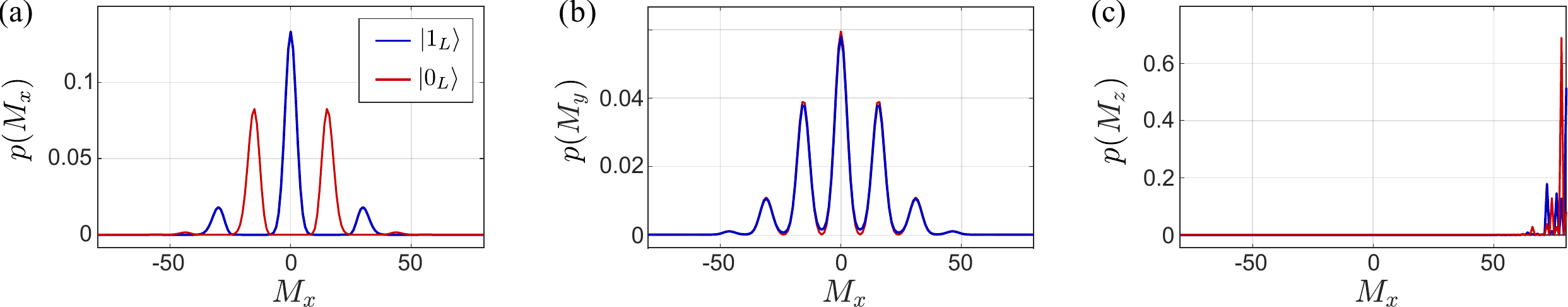}
    \caption{Collective-spin probability distributions for spin-GKP codewords, Eq.~\eqref{eqn:spingkpdef}, with $N=160$, $\delta=0.4$, and $\mathbf{T}=(5,5)$.
    (a) $J_x$-basis.
    (b) $J_y$-basis; note here that the distributions are nearly identical.
    (c) $J_z$-basis highlighting concentration near $\frac{N}{2}$, indicative of the HP regime. For the states shown, $\bra{\logic{0}} \hat{J}_z \ket{\logic{0}} = \tfrac{N}{2}-2.65$ and $\bra{\logic{1}} \hat{J}_z \ket{\logic{1}} = \tfrac{N}{2}-3$.
    }
    \label{fig:prob_spin_gkp_j_160}
\end{figure*}

\section{spin-GKP codes}

The first family of HP spin codes, spin-GKP codes, were introduced by Omanakuttan and Volkoff in Ref.~\cite{spin_GKP} to import single-mode square-lattice GKP codes into spin systems.
They are approximate quantum error-correcting codes that, for a single large-$J$ spin, outperform other known spin codes for the rotation-error class $\mathcal{E} = \{\hat{I}, \hat{J}_x, \hat{J}_y, \hat{J}_z\}$ and admit an explicit error-correction protocol.  In the spin-ensemble setting, Pauli operations and an entangling Clifford arise from simple powers of collective spin operators, and Pauli measurements are collective spin projections. When supplemented with magic states, the construction becomes universal and fault-tolerant. Here, we summarize spin-GKP codes and provide new details relative to the HP spin-code framework introduced here, as we emphasize spin-GKP codes in later sections.

The CV-GKP stabilizers,
    \begin{subequations}
        \begin{align}
        \hat{S}_X &\coloneqq e^{-i 2\sqrt{\pi}\, \hat{p}} = \hat{D} \big(  \sqrt{2\pi} \big),
        \\
        \hat{S}_Z &\coloneqq e^{+i 2\sqrt{\pi}\, \hat{q}} = \hat{D} \big( i\sqrt{2\pi} \big),
        \end{align}
    \end{subequations}
describe orthogonal displacements in phase space, and they commute $[\hat{S}_X,\hat{S}_Z] = 0$. 
Using the correspondence in Eq.~\eqref{eq:disptorot}, the CV GKP stabilizers are mapped to small rotations on a spin 
    \begin{subequations} \label{eq:stabs_spinGKP}
        \begin{align}
        \hat{S}_X &\mapsto \hat{T}_X \coloneqq 
    e^{-i\, 2\sqrt{\tfrac{2\pi}{N}}\, \hat{J}_y}
    = \hat{R} \Big( 2\sqrt{\tfrac{2\pi}{N}},\, 0 \Big),
    \label{eq:stabilizer_spin_gkp_x}
    \\
    \hat{S}_Z &\mapsto \hat{T}_Z \coloneqq e^{+i\, 2\sqrt{\tfrac{2\pi}{N}}\, \hat{J}_x}
    = \hat{R} \Big(2\sqrt{\tfrac{2\pi}{N}},\, \tfrac{\pi}{2} \Big),
\label{eq:stabilizers_spin_gkp_z}
        \end{align}
    \end{subequations}
where $\hat{R}(\SCSangle)$ is defined in Eq.~\eqref{eq:SU2displacement}. Combining the $\mathrm{SU}(2)$-displacements,
\begin{align}
    &\hat{T}_X \hat{T}_Z \approx \hat{R} \Big(2\sqrt{2}\sqrt{\tfrac{2\pi}{N}},\, \tfrac{\pi}{4}\Big)\, e^{-i\, 2 \tan^{-1} \left(\tfrac{2\pi}{N}\right)\hat{J}_z},\\
    &\hat{T}_Z \hat{T}_X \approx \hat{R} \Big(2\sqrt{2}\sqrt{\tfrac{2\pi}{N}},\, \tfrac{\pi}{4}\Big)\, e^{+i\, 2 \tan^{-1} \left(\tfrac{2\pi}{N}\right)\hat{J}_z},
\end{align}
reveals that a residual $z$-rotation whose angle scales as $\sim \frac{1}{N}$. Consequently, the commutator
\begin{equation}
    [\hat{T}_X,\hat{T}_Z]
    \approx
    2i\, \hat{R} \Big(2\sqrt{2}\sqrt{\tfrac{2\pi}{N}},\, \tfrac{\pi}{4}\Big)\,
    \sin \Big(2 \tan^{-1}\!\tfrac{2\pi}{N}\, \hat{J}_z\Big),
\end{equation}
vanishes as $N\to\infty$, recovering the desired commutation relation. 

Spin-GKP Pauli gates are small rotations,
\begin{equation}
    \hat{X} = e^{-i \sqrt{\tfrac{2\pi}{N}}\, \hat{J}_y},\qquad
    \hat{Z} = e^{-i \sqrt{\tfrac{2\pi}{N}}\, \hat{J}_x},
\end{equation}
and a complete set of Clifford gates is
    \begin{equation} \label{eq:GKPCliffords}
        \hat{H} = e^{i \frac{\pi}{2}\, \hat{J}_z}, \quad
        \hat{S} = e^{i \frac{1}{N}\, \hat{J}_x^2}, \quad
        \text{CNOT} = e^{-i \frac{2}{N}\, \hat{J}_x \otimes \hat{J}_y}.
    \end{equation}
Non-destructive Pauli and stabilizer measurements use ancillae prepared in $\ket{\logic{0}}$ or $\ket{\logic{+}}$, which are coupled to data via CNOTs and followed by measurements in the $\hat J_x$ or $\hat J_y$ basis. Outcomes are decoded by binning to the nearest multiple of $\sqrt{2\pi/N}$, in direct analogy with standard CV-GKP syndrome decoding.
Since the unitary $\hat{S}$ above spreads errors non-linearly~\cite{Omanakuttan_spincats_PRXQ_2024}, one may prefer a teleported, fault-tolerant implementation using the $Y$-eigenstate $\ket{\logic{i}}$. A convenient universal logical set is then
    \begin{equation}
        \bigl\{ \ket{\logic{0}},\, \mathcal{M}_Z,\, \hat{H},\, \text{CNOT} \bigr\}
        \;\cup\;
        \bigl\{ \ket{\logic{i}},\, \ket{\logic{T}} \bigr\}.
        \label{eq:logical_level_gates}
    \end{equation}
Preparation of the magic state $\ket{\logic{T}}$ can follow the CV-GKP recipe in Ref.~\cite{Ben_GKP_distillation}: start from a easy-to-prepare, non-encoded state such as the fully polarized state $\ket{\frac{N}{2},\frac{N}{2}}$, perform error correction, and obtain with high probability a state with sufficient magic for distillation. 
See Appendix~\ref{sec:magic_states_spin_GKP} for a sketch of this procedure.

\begin{figure*}
    \centering
    \includegraphics[width=0.96\linewidth]{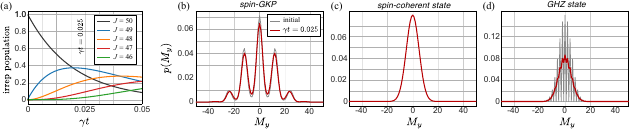}
    \caption{
Effects of local-symmetric depolarizing noise on states of $N=100$ spins. For a spin-GKP $\ket{\logic{0}}$ state, (a) shows the total population in each spin-$J$ irrep after time $\gamma t$. Probability distributions for collective measurements in the $\hat{J}_y$ basis given (b) this spin-GKP state, (c) a SCS along $z$, and (d) the GHZ state in Eq.~\eqref{eq:GHZ}, respectively. Shown are initial (grey) and partially depolarized (red) distributions for $\gamma t = 0.025$, which is marked on subplot (a) by a dashed line. 
}
    \label{fig:irreps+dist}
\end{figure*}

What is the structure of spin-GKP codewords? Ideal CV-GKP states have infinite energy (they are infinitely sharp grid states), and finite-energy approximations to them are not unique~\cite{GKP_Gottesman_PRA_2001,PhysRevA.97.032346,matsuura_GKPequiv}. The same is true for spin-GKP at finite $N$, and Ref.~\cite{spin_GKP} defines several different classes of spin-GKP codes depending on their construction. Such classes have finite-energy analogues in the CV domain, all of which converge in the limit of infinite energy, corresponding to $N \rightarrow \infty$. Here, we focus on the class \texttt{spingkp} with computational basis codewords
\begin{equation}
    \label{eqn:spingkpdef}
    \ket{\logic \cwlabel}
    = \sum_{\mathbf{t} = -\mathbf{T}}^{\mathbf{T}}
      \beta_{\mathbf{t}}\,
      e^{-i(2 t_1+\mu)\sqrt{\tfrac{2\pi}{N}}\, \hat{J}_y}\,
      e^{\, i t_2 \sqrt{\tfrac{2\pi}{N}}\, \hat{J}_x}\,
      \ket{\frac{N}{2},\, \frac{N}{2}},
\end{equation}
where $\mathbf{T} \coloneqq (t_1^\text{max}, t_2^\text{max})$ is a lattice truncation, and coefficients $\beta_{\mathbf{n}}$ mimic the Gaussian envelope of damped CV-GKP codes via the de Moivre–Laplace approximation,
    \begin{equation}
    \beta_{\mathbf{t}}
    \;\propto\;
    \frac{\Gamma(N+1)}
         {\Gamma \big(\tfrac{N}{2}+g_{\mathbf{t},\delta}+1\big)\,
          \Gamma \big(\tfrac{N}{2}-g_{\mathbf{t},\delta}+1\big)}
          \sim 
           \sqrt{\tfrac{2}{\pi N}}\, 
          e^{-\tfrac{2 g_{\mathbf{t},\delta}^2}{N} } ,
    \end{equation}
where $\Gamma(\cdot)$ is the gamma function, and
    \begin{equation}
        g_{\mathbf{t},\delta}
        \coloneqq \frac{\delta \sqrt{N\pi}}{2}\,
              \sqrt{(2t_1+\mu)^2 + t_2^2}.
    \end{equation}
The free parameter $\delta$ plays the role of a an energy damping parameter and can be tuned to improve orthogonality. An example spin-GKP codeword for $N=160$ and $\delta=0.4$ is shown in \cref{fig:prob_spin_gkp_j_160}.

The lattice truncation $\mathbf{T}$ and the $\beta_\mathbf{t}$ work in tandem to ensure that the HP approximation is valid. This can be seen by combining the $\mathrm{SU}(2)$ displacements to give an alternate description of the codewords in terms of SCSs,
    \begin{equation}
        \label{eqn:spingkpdef1}
        \ket{\logic \cwlabel}
        = \sum_{\mathbf{t} = -\mathbf{T}}^{\mathbf{T}} \beta_{\mathbf{t}}\,
      \ket{ \SCSangle_\mathbf{t} },
    \end{equation}
whose SCS angles for $N \gg 1$ are given by,
\begin{align}
    \theta_{\mathbf{t}}
    &= \frac{2}{\sqrt{N}} \,
       \sqrt{\Big(\sqrt{2\pi}\, t_1 + \mu \sqrt{\tfrac{\pi}{2}}\Big)^2 + \frac{\pi}{2}\, t_2^2},
    \\
    \phi_{\mathbf{t}}
    &= \tan^{-1} \left(
        \frac{\sqrt{\tfrac{\pi}{2}}\, t_2}
             {\sqrt{2\pi}\, t_{1} + \mu \sqrt{\tfrac{\pi}{2}}}
      \right) + \mathcal{O} \left( \frac{1}{N} \right).
\end{align}
$\mathbf{T}$ bounds the spin-lattice points (SCSs locations) away from $\theta = 0$, and the coefficients $\beta_\mathbf{t}$ smoothly damp SCS-components such that the weight of the lattice superposition is exponentially suppressed near the truncation. In principle $\beta_\mathbf{n}$ can provide this damping in the absence of a lattice cutoff; including $\mathbf{T}$ gives more freedom to parameterize the states, particularly for simulations. Together, these choices ensure the HP condition that the $Q_\text{spin}$ function remains localized within the region of spin phase space near the north pole that shrinks as $N^{-1/2}$.

\section{Local noise in the Holstein-Primakoff Approximation}
\label{sec:Local_noise_in_HPA}

\begin{figure*}
    \centering
    \includegraphics[width=0.96\linewidth]{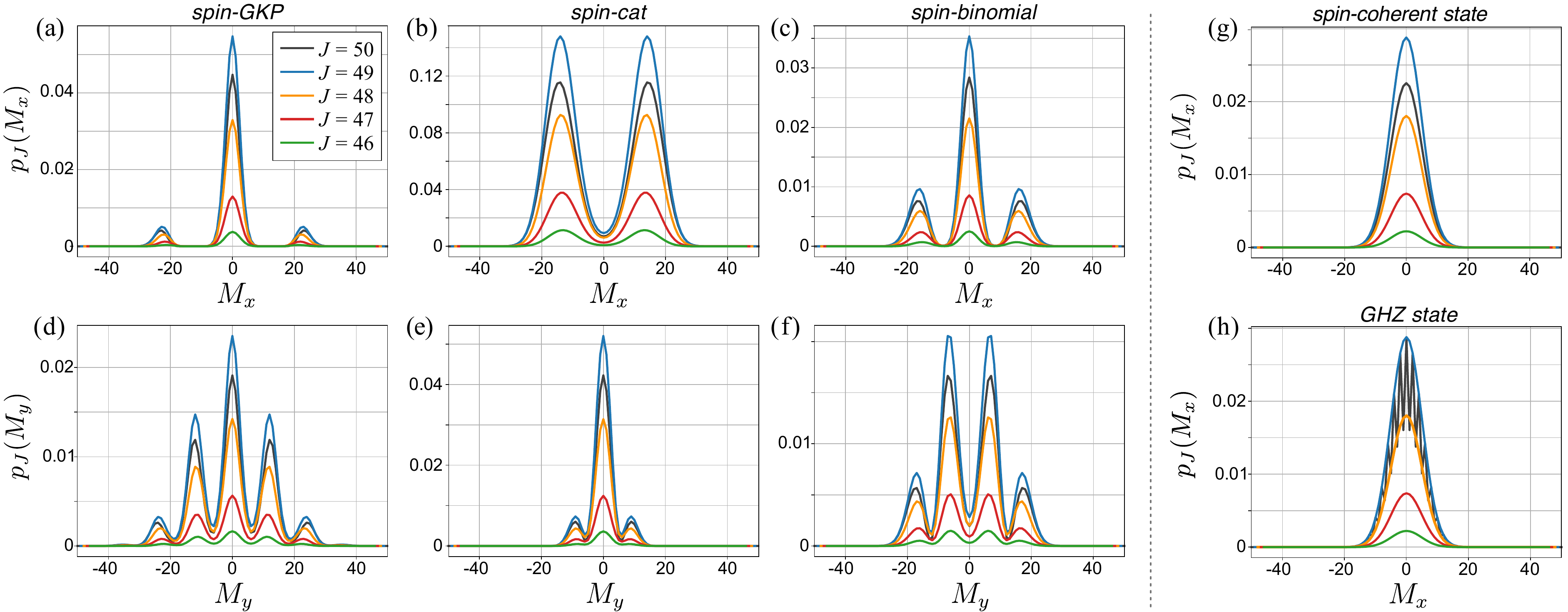}
    \caption{
Self-similarity of HP spin-code states across spin-irreps under local-symmetric depolarizing noise. 
Probability distributions in each $J$-irrep for  transverse collective spin observables
$\hat J_x$ (a--c) and $\hat J_y$ (d--f) for three representative HP spin codes for $N=100$ spins:
a spin-GKP code, a spin-cat code, and the smallest spin-binomial code. Each code is prepared in $\ket{\logic{0}}$ with parameters tuned such that $\expt{\hat{J}_z} = J-2 = 48$. 
The state evolves under the local symmetric depolarizing channel, Eq.~\eqref{eq:localsymmetricnoise}, with noise strength $\gamma t = 0.025$, after which it is projected into various spin-irreps, and the corresponding marginal distributions are found. 
Note that at $\gamma t = 0.025$, the total probability is highest in the $J=49$ irrep, see Fig.~\ref{fig:irreps+dist}(a), which is why the orange curves are above blue. For comparison, we also show in (g) a SCS along $z$ and in (h) a GHZ state, Eq.~\eqref{eq:GHZ}. While the SCS displays self-similarity as it is the canonical HPA state, the GHZ (a non-HP state) does not---the features arising from interference fringes that are present in the $J=50$ irrep are entirely washed out in lower irreps. 
}
    \label{fig:codewords_in_irreps}
\end{figure*}

The KL conditions above indicate that the logical effects of local noise are suppressed for HP spin codes. Here, we investigate in detail how local noise processes affect states prepared in the symmetric subspace that satisfy the HP approximation.  We focus primarily on local depolarizing noise, although the analysis generalizes to other local noise processes. Local depolarizing noise is modeled using a Lindblad master equation with jump operators $\hat{j}_i = \frac{1}{2} \hat{\sigma}_i$ for $i = x,y,z$ applied at rates $\gamma_i = \gamma$, 
    \begin{equation}
        \frac{d \hat \rho}{d t} = \frac{\gamma}{4} \sum_{i=x,y,z}^3(\hat \sigma_i \hat \rho \hat \sigma_i - \hat \rho).
    \end{equation}
For $N$ spins, consider the case where the depolarizing rates are equal across all spins such that the noise acts symmetrically. This yields the master equation
    \begin{equation} \label{eq:localsymmetricnoise}
        \frac{d \hat \rho}{d t} \coloneqq 
        \frac{\gamma }{4}
        \bigg[ \sum_{n=1}^N \sum_{i=x,y,z}
        \hat{\sigma}^{(n)}_{i} \hat\rho \big(\hat{\sigma}^{(n)}_i \big)^{\dagger} \bigg]
        -
        \gamma  \frac{3 N}{4} \hat\rho \, .
    \end{equation}
Given an initial collective state $\bar{\rho}$, Eq.~\eqref{eq:collective_state}, it was shown that any local-symmetric master equation induces dynamics that couple block-diagonal $J$-irreps but do not introduce coherences between them~\cite{chase2008, Baragiola2010}. 

Figure~\ref{fig:irreps+dist}(a), shows how local depolarizing noise transfers population from the symmetric subspace to other irreps for an initial spin-GKP state. In spite of significant pumping between irreps, the probability distribution for collective spin measurements is not significantly affected, as shown in Fig.~\ref{fig:irreps+dist}(b). For comparison, we show two other representative states: a SCS along $\hat{J}_z$, which is the ``canonical" HP state, and a GHZ state, which does not satisfy the HP approximation. The distribution for a SCS is unchanged by the dynamics, because depolarizing noise does affect transverse spin distributions for a SCS; even the extreme ase of a maximally mixed state has the same distribution. In contrast, GHZ states are extremely sensitive to noise, and we see in Fig.~\ref{fig:irreps+dist}(d) that the spikes in the initial distribution are almost entirely washed out.

The resilience of spin-GKP codes is tied to the fact that they exhibit \emph{self-similarity} across irreps as the states are damaged by local noise. This is not unique to spin-GKP; it is a feature of all HP spin codes.  In Fig.~\ref{fig:codewords_in_irreps}, we consider three representative HP spin codes imported from bosonic codes---spin-GKP, a two-lobe spin-cat code, and the smallest spin-binomial code. Each code is initialized in $\ket{ \logic 0} $ and then evolves under Eq.~\eqref{eq:localsymmetricnoise} for  $\gamma t = 0.025$.  Then, we separate the resulting state into spin irreps, $J \in \{\frac{N}{2}, \frac{N}{2} - 1, \dots \}$, using the projectors,
    \begin{equation} \label{eq:irrepprojection}
        \hat{P}_J \coloneqq \sum_{M = -J}^J \sum_{\lambda = 1}^{d^J_N} \ketbra{J, M, \lambda}{J,M, \lambda} ,
    \end{equation}
and compute the corresponding unnormalized probability distributions for the transverse spin observables within each irrep, $p_J(M_x)$ and $p_J(M_y)$. A striking feature is the self-similar structure of the distributions across irreps in Fig.~\ref{fig:codewords_in_irreps}(a-f). That is, although population leaks from the symmetric subspace into other irreps, the functional form of the distributions remains: grid-like peaks for the spin-GKP state, the interference-fringe patterns for the spin-cat state in $\hat{J}_y$, and the multi-lobed structure of the binomial code.  
The self-similarity indicates that the noise has not significantly damaged the code words as it mixes them across irreps, which agrees with the conclusion from the HP spin-code KL conditions that local noise does little logical damage.

This self-similarity is a direct consequence of the HP approximation. We saw in Fig.~\ref{fig:irreps+dist} a SCS fully polarized along $\hat{J}_z$, exhibits no change at all in $p(M_i)$. This is confirmed by Fig.~\ref{fig:codewords_in_irreps}(g), where the distributions in the lower irreps are identical in shape to that in the symmetric subspace and reproduce the full distribution via $p(M_i) = \sum_J p_J(M_i)$.
In contrast, the GHZ state, which cannot satisfy an HP approximation along any axis, is shown in Fig.~\ref{fig:codewords_in_irreps}(h).  The sharp interference fringes persist only in the symmetric subspace and are completely washed out in the other irreps. In the sum over irreps, only a trace of these features remains, see Fig.~\ref{fig:irreps+dist}(d).

\subsection{Short-time dynamics}

To further understand the self-similarity in Fig.~\ref{fig:codewords_in_irreps}, we examine the non-trivial short-term dynamics induced by the master equation, Eq.~\eqref{eq:localsymmetricnoise}, interpreted as an ensemble average over stochastic maps describing two processes: the ``jump'' part of the evolution describing a depolarizing event, and the ``no jump'' part describing simple decay of the norm of the collective state. We consider an initial PI state $\bar \rho_0$ in the symmetric subspace, with the overbar indicating collective states, Eq.~\eqref{eq:collective_state}. Initially, the no-jump part of the evolution does not affect $\bar \rho_0$ until a depolarizing jump occurs. At that point, the unnormalized state is
    \begin{equation} \label{eq:jumpmap}
        \bar{\rho}_\text{jump} 
        = 
        \frac{\gamma }{4}
        \sum_{n=1}^N \sum_{i}
        \hat{\sigma}^{(n)}_i \bar{\rho}_0 \big(\hat{\sigma}^{(n)}_i \big)^{\dagger} .
    \end{equation}
Appendix~\ref{appendix:symdepolarizingME}, we show that a single, symmetrized depolarizing event drives $\bar \rho_0$ into a mixture over the two largest $J$-irreps, which can be expressed as
    \begin{equation} \label{eq:jumpstate}
        \bar{\rho}_\text{jump} = \bar{\rho}^{(\frac{N}{2})}_\text{jump} \oplus \bar{\rho}^{(\frac{N}{2}-1)}_\text{jump},
    \end{equation}
where $\bar{\rho}^{(J)} \coloneqq \hat{P}_J \bar{\rho} \hat{P}_J$ is the unnormalized portion of the state projected into irrep $J$. Expressing the initial state in terms of SCSs, 
    \begin{equation} \label{eq:stateinSCSbasis}
        \bar{\rho} = \sum_{k, k'} \beta_k \beta^*_{k'} \ketbra{\SCSangle_k}{\SCSangle_{k'}},
    \end{equation}
the irrep projections in Eq.~\eqref{eq:jumpstate}, evaluated in Appendix~\ref{appendix:symdepolarizingME}, are
    \begin{subequations} \label{eq:jumpstateirreps}
    \begin{align}
        &\bar{\rho}^{(\frac{N}{2})}_\text{jump}  =
        \frac{\gamma}{N}\sum_{i=x,y,z} \hat J_i \bar\rho_0 \hat J_i,
        \label{eq:jumpstateirreps_0}
        \\ 
        &\bar{\rho}^{(\frac{N}{2}-1)}_\text{jump} 
        =  \label{eq:final_state_symmetric_local_noise_main}
        \\
        & \frac{\gamma (N-1)}{4} 
        \sum_{k,k'}\beta_k \beta^*_{k'} 
        \cos^2 \left(\frac{\Delta \SCSangle_{k,k'}}{2} \right)
        \overline{\ketbra{\SCSangle_k}{\SCSangle_{k'}}}^{(N/2-1)}, 
        \nonumber
    \end{align}
    \end{subequations}
where $\Delta \SCSangle_{k,k'}$ is the total angular separation between SCSs, Eq.~\eqref{eq:angularseparation}, and 
    \begin{equation} \label{eq:collectivestateSCSop}
        \overline{\ketbra{\SCSangle_k}{\SCSangle_{k'}}}^{(J)} 
        \coloneqq
        \hat R(\SCSangle_k) \overline{\ketbra{J,J}} \hat R^\dagger(\SCSangle_{k'})
    \end{equation}
describes SCS coherences in irrep $J$. Thus, Eq.~\eqref{eq:final_state_symmetric_local_noise_main} is a collective state.

The irrep projections in Eq.~\eqref{eq:jumpstateirreps} apply for any initial state $\bar \rho_0$ in the symmetric subspace. Expanding the geometric cosine factor, Eq.~\eqref{eq:final_state_symmetric_local_noise_main},
    \begin{equation}
        \cos^2 \left(\frac{\Delta\SCSangle_{k,k'}}{2}\right)
        =
        1 - \frac{\Delta\SCSangle_{k,k'}^{\,2}}{4}
        + \mathcal{O} \left( \Delta\SCSangle_{k,k'}^{\,4} \right),
    \label{eq:cosSquaredExpansion}
\end{equation}
we see that the leading term reproduces the SCS superposition in the initial state, Eq.~\eqref{eq:stateinSCSbasis}, here localized to the $J-1$ irrep. 
States imported via the HP approximation have polar angle scaling $\theta_k = \mathcal{O}(N^{-1/2})$, Eq.~\eqref{eq:SCScoeffs}, and thus $\Delta \SCSangle$, Eq.~\eqref{eq:angularseparation}, has the same scaling.
Therefore, we obtain
    \begin{equation}
        \bar{\rho}^{(\frac{N}{2}-1)}_\text{jump} 
        = \frac{\gamma(N-1) }{4} \Big[ \bar{\rho}_0^{(\frac{N}{2}-1)} + \mathcal{O}(N^{-1}) \Big] ,
        \label{eq:self_similar_local_symmetric}
    \end{equation}
demonstrating that the projection onto the $J=\frac{N}{2}-1$ irrep preserves the structure of an HP state for large $N$. We find that the total state, Eq.~\eqref{eq:jumpstate}, additionally experiences a dephasing-type map in the symmetric subspace. However, this process occurs at a relative rate two orders of magnitude smaller in $N$, and the dominant process is the self-similar mixing over irreps. 

Outside of the HP approximation, $\Delta \SCSangle$ can be large due to large polar-angle separation, and the portion of the state in lower irreps will differ significantly from the portion in the symmetric subspace. An example is the GHZ state with $\Delta \SCSangle = \pi$, giving $\cos^2 (\Delta \SCSangle/2) = 0$. The ``self-similarity'' term in Eq.~\eqref{eq:final_state_symmetric_local_noise_main} vanishes entirely, as seen in Fig.~\eqref{fig:codewords_in_irreps}(h).

\subsection{Asymmetric local noise}

\begin{figure}
    \centering
     {\includegraphics[width =0.9 \columnwidth]{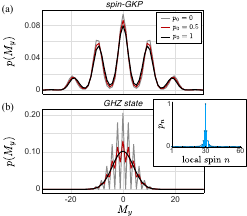}} 
    \caption{
    Effects of a spin-dependent depolarizing map, Eq.~\eqref{eq:asym_noise_profile} with $\zeta = 1$ and varying $p_0$ over $N=60$ spins. The distribution of $p_n$ is shown in the inset plot for $p_0 = 1$. $\hat{J}_y$ probability distributions are shown in (a) for a spin-GKP $\ket{\logic{0}}$ and in (b) for a GHZ state. 
  }
    \label{fig:tensor_simulation}
\end{figure}

The preceding analysis focused on permutation-symmetric local noise, where the dynamics can be understood entirely in terms of collective operators and inter-irrep mixing. In realistic settings, however, local error rates may vary across the spins, explicitly breaking permutation invariance. Nevertheless, the KL conditions do not rely on permutation invariance, and HP spin codes remain resilient to inhomogeneous local noise. 

To show this in practice, consider an inhomogeneous local depolarizing map where the $n$th spin has a probability to undergo a depolarizing event with probability $p_n$ that varies across the spin ensemble,
    \begin{equation}
        p_n = p_{0}\exp ( -\zeta | n - \tfrac{N}{2} | ).
        \label{eq:asym_noise_profile}
     \end{equation}
Large $\zeta$ corresponds to strongly localized noise acting on a small subset of spins, while smaller $\zeta$ approaches the symmetrized-noise limit.
We simulate this error model using a tensor-network–based algorithm previously developed to study spatially correlated noise on PI states~\cite{Tylertensornetwork}. We first construct a matrix product state (MPS) representation~\cite{Orus_2014, Bridgeman_2017, Cirac_2021} of initial pure states. Since these states lie in the symmetric subspace, their bond dimension scales at most linearly with $N$~\cite{perez-garcia_2007}. The corresponding matrix product operator (MPO) then has bond dimension at most quadratic in $N$, and its evolution under the inhomogenous depolarizing map does not increase the bond dimension, as this channel is uncorrelated across sites. Similarly, MPO representations of projectors onto collective spin eigenspaces are constructed with bond dimension linear in $N$, allowing efficient computation of outcome probabilities.
In \cref{fig:tensor_simulation}, we show the transverse probability distributions for a spin-GKP and a GHZ state undergoing a highly localized inhomogenous depolarizing map. 
For spin-GKP, we observe the same self-similarity as for local-symmetric noise: the structure of the probability distribution is maintained even as some spins become fully depolarized. 
In contrast, the fragility of the GHZ state is apparent in the rapid loss of fringes in the probability distribution.

\section{Measurement-free local error recovery}
\label{sec:measurement_free_error_correction}

A central challenge in quantum error correction is the reliance on syndrome measurements, decoding, and corrections, which introduce latency and additional noise. For local-spin errors that drive the state out of the symmetric subspace, the HP spin-code framework naturally supports measurement-free local error recovery (MFLER), where the state is returned to the symmetric subspace coherently without intermediate measurements or feed-forward. MFLER is not a replacement for logical QEC but rather serves as a coherent leakage-recovery step that converts local errors into collective errors, which can then be dealt with by the HP spin code.

An HP spin-code state $\ket{\logic{\psi}}$ that has been damaged by local-symmetric noise has nontrivial support over other spin irreps, $\ketbra{\logic \psi}{ \logic \psi} \rightarrow \bigoplus_J \bar{\rho}^{(J)}$. Two important properties of the damaged state when the noise is not too strong: (\emph{i}) only irreps near $J_\text{max} = \frac{N}{2}$ are significantly populated, and (\emph{ii}) the $\bar{\rho}^{(J)}$ satisfy the HP approximation.
The MFLER procedure uses a circuit to swap the damaged data ensemble into the symmetric subspace using an ancilla ensemble initialized in $\ket{\logic{+}}$. For each $\bar{\rho}^{(J)}$ it functions as
    \begin{align}
        \begin{split} \label{eq:SWAPgadget}
        \Qcircuit @C=1.3em @R=1.6em {
            \lstick{\bar{\rho}^{(J)}} & \targ      & \ctrl{1} & \qw & \rstick{\ket{\logic +}^{(J)}} \\
            \lstick{\ket{\logic +}^{(J_\text{max})}}    & \ctrl{-1}  & \targ    & \qw & \rstick{\bar{\rho}^{(J_\text{max})}}
        } 
            \end{split}
    \end{align}
where, crucially, the CNOT gates are assumed to have the same logical action across the irreps. Additionally, this action assumes the self-similarity discussed above; \emph{i.e.} that $\bar{\rho}^{(J)}$ is approximately supported on the code space that one would obtain by importing a bosonic codes directly to the irrep with total spin $J$. The key property is that the states are swapped between the two ensembles without altering their total spin. Since the CNOT gates act across irreps, the SWAP acts on all the damaged portions of the state, simultaneously transferring every $\bar{\rho}^{(J)}$ into to the symmetric subspace and effectively converting local errors into collective errors that can later be dealt with using the QEC properties the HP spin code. 

The SWAP gadget in Eq.~\eqref{eq:SWAPgadget} is compatible in principle with any PI spin code, but its practical realization requires logical gates that act uniformly across spin irreps populated by local noise. For imported HP spin codes, this condition is naturally satisfied: logical CNOT gates are inherited from their bosonic counterparts via the HP approximation, and thus are generated by collective spin operators. More concretely, the canonical bosonic operators relate to their spin counterparts via Eqs.~\eqref{eq:HPAconds}, and a bosonic unitary $ \hat U_{\mathrm{CV}} = f(\hat a,\hat a^{\dagger})$ is therefore represented in the spin ensemble as $\hat U_{\mathrm{HP}} = f( J^{-1/2} \hat{J}_x, J^{-1/2} \hat{J}_y)$ with the same functional dependence as in the bosonic description.  We assume that the $J$-dependence of $\hat U_{\textrm{HP}}$ enters only through the arguments of $f$, \emph{i.e.} that the bosonic operator $f(\hat a,\hat a^{\dag})$ does not depend on $J$.  For irreps near the symmetric subspace with $J = \frac{N}{2} - \Delta J$ and $\Delta J = \mathcal O (1)$, the HP–scaled generators,
    \begin{equation}
        \frac{\hat J_{i}}{\sqrt{J}}
        = \frac{\hat J_{i}}{\sqrt{N/2}}
       \Big[ 1 + \mathcal{O} ( N^{-1} ) \Big] \quad \text{for} \quad i=x,y \, ,
    \end{equation}
differ from their action in the symmetric subspace only by relative $\mathcal O(N^{-1})$ factors. Each monomial in $f$ therefore incurs only the same $1 + \mathcal O(N^{-1})$ correction upon restriction to any irrep with $J = \frac{N}{2} - \mathcal O(1)$. Thus, when local noise redistributes weight only to irreps with $J=J_{\max} - \mathcal O(1)$, the HP approximation ensures that deviations in the logical action are suppressed as $\mathcal O(N^{-1})$. In this regime, the MFLER protocol coherently returns the state to the symmetric subspace, after which residual errors are handled by HP spin-code error correction. If substantial population accumulates on irreps far from $J_{\max}$, this uniformity is lost and the recovery protocol no longer faithfully restores the logical state.

We make this explicit for spin-GKP codes. The unitaries that enact spin-GKP Pauli and Clifford gates, Eq.~\eqref{eq:GKPCliffords}, act uniformly across all $\mathrm{SU}(2)$ irreps, since they are generated by collective spin operators. However, the intended logical action requires that $\hat{J}_x$ and $\hat{J}_y$ act like quadrature operators in the HP approximation, and the gate strengths are functions of total spin $J$. The spin-GKP entangling gate, see Eq.~\eqref{eq:GKPCliffords}, has an interaction strength $J^{-1}_\text{max} = \frac{2}{N}$ tuned to the largest $J$-irrep. The SWAP gadget in Eq.~\eqref{eq:SWAPgadget} relies on this CNOT gate having the same logical action across other irreps where the state has support. Consider a spin-GKP CNOT gate tuned to act across two irreps with respective total spins $J_1$ and $J_2$, $ \text{CNOT}_{J_1,J_2} \coloneqq \text{exp}[ -i (J_1 J_2)^{-1/2} \hat{J}_x \otimes \hat{J}_y ] $. Expressing $J_1 = \frac{N}{2} - \Delta J_1 $ and $J_2 = \frac{N}{2} - \Delta J_2$ and working in the limit ($\Delta J_1, \Delta J_2) = \mathcal{O}(1)$, we get $\text{CNOT}_{J_1,J_2} = \text{CNOT} + \mathcal{O}(N^{-1})$. Thus, for irreps with total spin $J$ close to $\frac{N}{2}$, the spin-GKP CNOT gate acts almost identically to its action within the fully symmetric subspace. 

\begin{figure}
\includegraphics[width = 0.85\columnwidth]{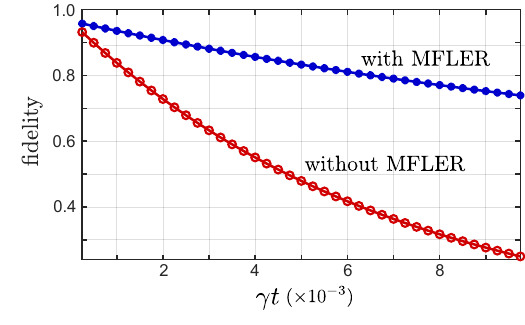} 
\caption{ Idealized MFLER for spin-GKP codes.
Fidelity with the initial spin-GKP state $\ket{ \logic{0} } $ for $N = 60$ after it undergoes symmetric local depolarizing noise of strength $\gamma t$. The red curve shows fidelity to the initial state without MFLER, and the blue curve shows the fidelity after MFLER that transfers population in the damaged state from lower irreps back to the symmetric subspace using the SWAP gadget in \cref{eq:SWAPgadget}.
}
\label{fig:recovery_fidelity}
\end{figure}

The performance of MFLER is shown for spin-GKP codes in \cref{fig:recovery_fidelity}. Due to simulation constraints, we were not able to simulate the full circuit. Rather, we assume a proof-of-principle, idealized MFLER protocol that perfectly transfers the population from lower irreps into the symmetric subspace. Due to self-similarity under local depolarizing noise, the majority of the damage to the state at short times comes from irrep-population redistribution, see Fig.~\ref{fig:irreps+dist}(a). The red curve shows the fidelity after the noise, and the blue curve includes the idealized MFLER operation that repopulates the symmetric subspace.
Even at moderate $N$, the fidelity is substantially restored, with larger $N$ expected to perform better due to better satisfaction of the HP approximation and more self-similarity over irreps near $J_\text{max}$.

\subsection{Mapping local errors to collective errors for spin-GKP codes}
\label{subsec:effective-kraus}

The MFLER protocol in Eq.~\eqref{eq:SWAPgadget} converts local-spin errors that mix $J$ irreps into errors solely in the symmetric subspace, which is spanned by a basis of collective spin operator. Therefore, it maps local errors into collective errors. The correctability of these errors depends on the error-correcting properties of the HP spin code. Focusing on spin-GKP codes, we show that, to first order in $\frac{1}{\sqrt{N}}$, any local single-qubit Pauli error is converted into a correctable collective error via the MFLER protocol above. This also extends the analysis above for local-symmetric noise to the case of genuinely local errors.

Consider two ensembles of equal size $N$ labelled $D$ (data) and $A$ (ancilla). We study the effect of a single-spin Pauli error on the data ensemble, $\hat \sigma^{D}_i$ for $i \in \{x,y,z\}$, followed by the MFLER protocol in Eq.~\eqref{eq:SWAPgadget}, which swaps the damaged state into the symmetric subspace of the ancilla ensemble and then discards the data ensemble. Due to permutation symmetry in the codewords, we need not specify which spin undergoes the Pauli error.
The resulting channel, $\mathcal{E} = \sum_m \hat{K}^i_m \odot (\hat{K}^i_m)^\dagger$, admits Kraus operators
    \begin{equation}
        \hat{K}_m^i \coloneqq {}^{D}\!\!\bra{ m } \hat U_{\rm SWAP} \, \hat \sigma^{D}_i \ket{ \logic{+} }^{A}_{J_\text{max}} ,
    \end{equation}
where $\{ \ket{m}^{D} \}$ is an arbitrary orthonormal basis for the discarded data ensemble, and $\sum_m (\hat K_m^i)^\dagger \hat K^i_m = \hat{I}$. 

To evaluate this expression, first note that
   \begin{equation}
        \hat{U}_\text{SWAP} \ket{\logic +}^{A}_{J_\text{max}} =  \ket{\logic +}^{D}_{J} \hat{U}_\text{SWAP}^{D \rightarrow A} 
    \end{equation}
where $\hat{U}_\text{SWAP}^{D \rightarrow A}$ is a partial swap from the data ensemble to the ancilla, and we have assumed that this operator will be acting on a state of the data ensemble 
that lies in the code space. Then, we can write
   \begin{equation} \label{eq:KrausOp_almost}
        \hat{K}_m^i  
        ={}^{D}\!\!\bra{m} \hat{U}_\text{SWAP}  \hat \sigma^{D}_i \hat{U}_\text{SWAP}^\dagger \ket{\logic +}^{D}_J \hat{U}_\text{SWAP}^{D \rightarrow A}.
    \end{equation}
Now that the input state $\ket{\logic{+}}$ and the measurement bras $\bra{m}$ are on the same system (the data ensemble), we can evaluate the Kraus operators. 

To do so, we focus on spin-GKP codes whose CNOT gate, Eq.~\eqref{eq:GKPCliffords}, allows us to write the SWAP gadget as $\hat U_\text{SWAP} = e^{\hat A}e^{\hat B}$, with generators given by collective spin operators,
    \begin{equation}
        \hat A \coloneqq -\tfrac{2i}{N} \hat J_x^{D} \otimes \hat J_y^{A},\quad
        \hat B \coloneqq -\tfrac{2i}{N} \hat J_y^{D} \otimes \hat J_x^{A}.
    \end{equation}
Then, for any single-qubit Pauli $\hat\sigma^{D}_i$, we get
    \begin{align} \label{eq:BCHonsigma}
        \hat U_\text{SWAP} \hat\sigma^{D}_i \hat U_\text{SWAP}^\dagger
        =
        \hat\sigma^{D}_i + [\hat B,\hat \sigma^{D}_i] + [\hat A,\hat  \sigma^{D}_i] + \mathcal O (N^{-1}),
    \end{align}
by Baker-Campbell-Hausdorff, with the commutators given by
    \begin{align}
        [\hat{A},\hat\sigma^{D}_x] &= 0, &
        [\hat{B},\hat\sigma^{D}_x] &= -\tfrac{2}{N}\hat\sigma^{D}_z \!\otimes\!\hat{J}^{A}_x, \nonumber
        \\[0.4em]
        [\hat{A},\hat\sigma^{D}_y] &= \tfrac{2}{N}\hat\sigma^{D}_z\!\otimes\!\hat{J}^{A}_y, &
        [\hat{B},\hat\sigma^{D}_y] &= 0,
        \\[0.4em]
        [\hat{A},\hat\sigma^{D}_z] &= -\tfrac{2}{N} \hat\sigma^{D}_y\!\otimes\!\hat{J}^{A}_y, &
        [\hat{B},\hat\sigma^{D}_z] &= \tfrac{2}{N}\hat\sigma^{D}_x\!\otimes\!\hat{J}^{A}_x. \nonumber
    \end{align}
Taking the partial matrix element in Eq.~\eqref{eq:KrausOp_almost} using Eq.~\eqref{eq:BCHonsigma} and defining $\eta_m^i \coloneqq \bra{m} \hat{\sigma}_i \ket{\logic{+}} = \mathcal{O}(1)$ yields the Kraus operators
    \begin{align} \label{eq:effective-kraus}
        \hat K^i_{m} 
        =
        \begin{cases}
            \eta_m^x \hat{I} - \frac{2}{N}\eta_m^z \hat J_x  &  \text{for }i =x,
            \\[4pt]
            \eta_m^y \hat{I} + \frac{2}{N}\eta_m^z \hat J_y   &  \text{for }i = y,
            \\[4pt]
            \eta_m^z \hat{I} - \frac{2}{N} \eta_m^y \hat J_y
            + \frac{2}{N} \eta_m^x \hat J_x  & \text{for }i = z,
        \end{cases}
    \end{align}
up to $ \mathcal{O}(N^{-1})$ corrections. Each Kraus operator lies in the linear span of $\{ \hat I,\hat J_x, \hat J_y \}$ and therefore also in the span of $\{\hat I,\hat J_x,\hat J_y,\hat J_z\}$). This defines correctable algebra for spin-GKP codes~\cite{spin_GKP}.

\section{Conclusion}
\label{sec:conclusion}
In this work, we introduced a class of PI codes, which we call HP spin codes, and a general framework for translating bosonic codes into HP spin codes on ensembles of qubits. From their bosonic counterparts, HP spin codes inherit the ability to correct collective spin errors. Moreover, HP spin codes can also suppress local-spin errors and leakage. Beyond this theoretical insight, we develop a measurement-free recovery protocol based on collective CNOT gates that act similarly across irreps in the HP regime. This protocol converts local errors into collective errors without requiring syndrome measurements, and we demonstrated through numerical simulations that spin-GKP states preserve their structure under symmetric local and asymmetric local noise. Together, these results position spin-GKP codes as promising candidates for fault tolerance. 

Our work opens several avenues for further research. Future studies can explore the properties of HP spin codes beyond spin-GKP, characterize thresholds for fault tolerance under realistic noise models, and explore implementations in atomic ensembles and other platforms wthat feature collective interactions and measurements. Also, we observed that symmetric and asymmetric local noise preserve the structure of collective spin distributions for HP spin codes, whereas the same noise processes rapidly destroy this structure for highly entangled GHZ states. Since GHZ states are maximally entangled by several standard measures, this contrast suggests that the resilience of HP spin codes is tied to a restricted entanglement structure in the HP approximation as compared to the rest of PI Hilbert space. 

While we focus here on codes in the context of quantum computing, bosonic codes and other optical states have demonstrated sensing applications~\cite{terhalGKPsensor2017,hypercube2019,Deng2024,catsensing2025} that can be explored in PI spin systems via the HP spin-code mapping. In such settings, aspects of sensing performance may persist under local noise due to the self-similar structure of states in the HP regime. Even spin states outside the strict HP regime that are useful for sensing~\cite{Zurek2001,Chalopin2018,Ouyangsensing2022,sivametrology2025,tang2025}, such as GHZ-type states, may benefit from being embedded into the HP subspace of a much larger spin system, potentially offering robustness against some local errors. However, various no-go theorems~\cite{2012ElusiveHeisenberg,Fujiwara_2008,Sekatski2017quantummetrology,DD2017,Zhou_2018} constrain simultaneous high-precision sensing and generic protection from local noise. Any emergent advantages must rely on additional structure in the signal or noise, therefore detailed analysis is required.

Finally, superselection-rule (SSR) formulations of bosonic quantum information emphasize descriptions restricted to fixed total particle number, with explicit quantum phase references and the CV formalism recovered only as an approximate limit of a symmetry-preserving framework~\cite{Milman2023, Descamps2024, Descamps2025HeisenbergWeyl}. This perspective is conceptually related to the HP approximation employed here, where CV behavior emerges from a constrained Hilbert space that retains particle-number and permutation invariance. These connections raise natural questions about how PI spin constructions including GKP-like encodings extend beyond the strict CV regime.

\begin{acknowledgements}
The authors acknowledge fruitful discussions with Ivan Deutsch, Jack Davis, Jonathan Gross, P\'{e}rola Milman, Nicolas Menicucci, and Shubham Jain. We thank in particular Rafael Alexander for the initial idea that sparked this collaboration.
B.Q.B. acknowledges support from the Australian Research Council Centre of Excellence for Quantum Computation and Communication Technology (Project No.~CE170100012) and from the Japan Science and Technology Agency through the Ministry of Education, Culture, Sports, Science, and Technology Quantum Leap Flagship Program. 

\end{acknowledgements}

\appendix

\section{Leakage Errors}
\label{sec:leakage_error}

Consider the Kraus operators for leakage 
    \begin{equation} \label{eq:kraus_atom_loss_all}
        \hat{E}_a^{(n)} = \hat{I}^{\otimes (n-1)} \otimes \bra{a} \otimes \hat{I}^{\otimes (N-n)}, 
    \end{equation}
where $a \in \{0,1\}$, and $1 \leq n \leq N$ labels the qubit that is lost. These Kraus operators describes a channel $\mathcal{E}_\text{leakage} = \sum_{a =0,1} \hat{E}_a \odot \hat{E}^\dagger_a$ where qubit $n$ is traced out. To determine the conditions under which such errors are correctable, we invoke the KL conditions, Eq.~\eqref{eq:KLconditions}. Assuming locality of errors, cross-terms with $n \neq n'$ vanish.

The single-body terms satisfy
    \begin{subequations}
    \begin{align}
        \big( \hat{E}_0^{(n)} \big)^{\dagger} \hat{E}_0^{(n)} 
        &= \hat{I}^{\otimes (n-1)} \otimes \frac{\hat{I} + \hat \sigma_z^{(n)}}{2} \otimes \hat{I}^{\otimes (N-n)} .
        \\
        \big( \hat{E}_1^{(n)} \big)^{\dagger} \hat{E}_1^{(n)}
        &= \hat{I}^{\otimes (n-1)} \otimes \frac{\hat{I} - \hat \sigma_z^{(n)}}{2} \otimes \hat{I}^{\otimes (N-n)}, \\
       \big( \hat{E}_0^{(n)} \big)^{\dagger} \hat{E}_1^{(n)}
        &= \hat{I}^{\otimes (n-1)} \otimes \hat \sigma_{+}^{(n)} \otimes \hat{I}^{\otimes (N-n)}, \\
        \big( \hat{E}_1^{(n)} \big)^{\dagger} \hat{E}_0^{(n)}
        &= \hat{I}^{\otimes (n-1)} \otimes \hat \sigma_{-}^{(n)} \otimes \hat{I}^{\otimes (N-n)} .
    \end{align}
    \end{subequations}
Therefore, any code that can detect the errors $\{\hat{\sigma}_z, \hat{\sigma}_+, \hat{\sigma}_-\}$ can correct leakage errors. 
Since this set spans all single-qubit Pauli operators, such a code also detects all single-Pauli errors.

In the terminology of Ref.~\cite{Aydin_permutationally_invariant_2024_Quantum}, tracing out a single spin corresponds to a \emph{deletion error}, i.e., complete loss of the spin with known location.  Although this operation formally maps the code space
from $N$ to $N-1$ spins, one may equivalently restore a fixed-$N$ register by introducing a fresh ancilla and applying erasure recovery.  The analysis above therefore shows that any code satisfying the KL conditions for single-spin Pauli operators can also correct a single deletion error.

\section{Action of local noise on the symmetric subspace}\label{appendix:symdepolarizingME}

Assuming the an initial collective state $\bar{\rho}$, 
the master equation in Eq.~\eqref{eq:localsymmetricnoise} can be written as
    \begin{equation} \label{eq:ME_sphericalbasis}
        \frac{d \bar \rho}{dt}
        = 
        \frac{\gamma}{4} 
        \sum_{j,q = -1}^1  \hat{L}_{j,q}  \bar{\rho} \hat{L}^\dagger_{j,q}  - \gamma \frac{3 N}{4} \bar{\rho} \, ,
    \end{equation}
with jump operators $\hat{L}_{j,q}$, defined in Ref.~\cite{Forbes_Collective_States}, that arise from projecting local Pauli operators into the spherical tensor operator basis~\cite{klimov2008generalized} and symmetrizing over the spins. They act on the collective matrix elements as
    \begin{equation} 
        \hat L_{j,q}\overline{\ketbra{J,M}{J,M'}} \hat L^\dagger_{j,q} \propto \overline{\ketbra{J+j,M+q}{J+j,M'+q}},
    \end{equation}
changing the total spin and projection quantum numbers by at most 1, because the local Pauli operators are rank-1 spherical tensor operators. Equation~\eqref{eq:ME_sphericalbasis} is not fully equivalent to the master equation in Eq.~\eqref{eq:localsymmetricnoise}, as it acts only within the subspace of collective states. Nevertheless, since the initial state is assumed to be in the symmetric subspace (which includes all HP spin codes), the dynamics are restricted to the collective states~\cite{chase2008}.
Also note that, although we focus on depolarizing noise here, this formalism applies to any local-symmetric Lindblad; more details can be found in Forbes \emph{et al.}~\cite{Forbes_Collective_States}.

\subsection{$J = \frac{N}{2}$ irrep}

In the symmetric subspace $J = J_{\max} = \frac{N}{2}$, we use the local-spin form of the jump map from Eq.~\eqref{eq:jumpmap}. Projected into this irrep
    \begin{equation} \label{eq:jumpmap}
        \bar{\rho}^{(\frac{N}{2})}_\text{jump} 
        = \frac{\gamma}{4}
        \sum_{n=1}^N \sum_{i=x,y,z}
        \hat P_{N/2}\hat{\sigma}^{(n)}_i\hat P_{N/2} \bar{\rho}_0 \hat P_{N/2}\big(\hat{\sigma}^{(n}_i \big)^{\dagger}\hat P_{N/2} .
    \end{equation}
By permutation symmetry of $\hat P_{N/2}$, the operators $\hat P_{N/2} \hat\sigma_i^{(n)} \hat P_{N/2}$ are independent of the spin index $n$. Summing over $n$ and using Eq.~\eqref{eq:collective_operator} gives $\sum_{n=1}^N \hat P_{N/2} \hat\sigma_i^{(n)} \hat P_{N/2} = 2 \hat J_i^{(N/2)}$, where $\hat J_i^{(J)}$ is the collective spin operator acting only in irrep $J$. Since the left-hand side consists of $N$ identical terms, we obtain
    \begin{equation}
        \hat P_{N/2}\,\hat\sigma_i^{(n)}\,\hat P_{N/2}
        =
        \tfrac{2}{N}\,\hat J_i^{(N/2)} .\label{eq:pauli_projected}
    \end{equation}
Therefore, the map in the symmetric subspace,
    \begin{equation} 
        \bar{\rho}^{(\frac{N}{2})}_\text{jump} 
        = 
        \frac{\gamma}{N} \sum_{i=x,y,z}
        \hat J_i \bar{\rho}_0  \hat J_i  ,
    \end{equation}
is a \emph{collective} depolarizing-type map. Note that Eq.~\eqref{eq:pauli_projected} is valid for projections into any irrep, not just the symmetric subspace. Therefore, local depolarizing noise can be described as collective depolarization within each irrep with simultaneous population transfer to and from other irreps (assuming dynamics only within the subspace of collective states).

\subsection{$J = \frac{N}{2}-1$ irrep}
For the $\frac{N}{2}-1$ irrep, we expresse the input state in the SCS basis, Eq.~\eqref{eq:stateinSCSbasis}, and use the jump expression for the master equation in Eq.~\eqref{eq:ME_sphericalbasis},
    \begin{equation} \label{eq:jumptermA}
        \bar{\rho}_\text{jump} = 
        \frac{\gamma}{4} \sum_{k,k'}
         \sum_{j,q = -1}^1 \beta_k \beta^*_{k'} \hat{L}_{j,q}  \ketbra{\SCSangle_k}{\SCSangle_{k'}} \hat{L}^\dagger_{j,q}.
    \end{equation}
Consider a single $k,k'$-term in Eq.~\eqref{eq:jumptermA}, which isolates two SCSs $\ket{\SCSangle}$ and $\ket{\SCSangle'}$ where $\SCSangle=(\theta,\phi)$ and $\SCSangle'=(\theta',\phi')$ giving total angular separation $\Delta \SCSangle$, Eq.~\eqref{eq:angularseparation}.
By symmetry of the depolarizing channel, the map in Eq.~(\ref{eq:jumptermA}) must commute with arbitrary SU(2) rotations, and thus one can write
    \begin{align}
        \nonumber\sum_{j,q = -1}^1  &\hat{L}_{j,q} \ketbra{\SCSangle}{\SCSangle'} \hat{L}^\dagger_{j,q}
        \\
        &=\hat R(\SCSangle) \Bigg[\sum_{j,q = -1}^1  \hat{L}_{j,q} \ketbra{0,0}{\Delta\boldsymbol{\Omega}} \hat{L}^\dagger_{j,q} \Bigg] \hat R^\dagger(\SCSangle).\label{eq:rotated_jump_term}
    \end{align}
This rotation makes the analysis in the collective state space much simpler, and highlights that for the depolarizing channel, only the total angular distance $\Delta\boldsymbol{\Omega}$ between SCSs is relevant.

Since we are interested in how the map in Eq.~(\ref{eq:jumptermA}) transfers population to the $J=\frac{N}{2}-1$ irrep, we project Eq.~(\ref{eq:rotated_jump_term}) onto that subspace using $\hat P_{N/2-1}$. Doing so, only the $j=q=-1$ term survives, since all other terms either annihilate $\ket{0,0}$, or transfer population only within the symmetric subspace. We note that
    \begin{equation}
        \hat L_{-1,-1}=\sum_{J,M} \sqrt{B^N_{J,M}} \ketbra{J-1,M-1}{J,M},
    \end{equation}
with coefficients $B^N_{J,M}$ given in Ref.~\cite{Forbes_Collective_States} that evaluate to $B^N_{N/2,N/2} = N-1$, giving
    \begin{equation} \label{eq:Lon00}
        \hat L_{-1,-1}\ket{0,0} = \sqrt{N-1} \ket{0,0}^{(\frac{N}{2}-1)},
    \end{equation}
where $\ket{0,0}^{(\frac{N}{2}-1)} \coloneqq \ket{\tfrac{N}{2}-1, \tfrac{N}{2}-1}$ is the maximally polarized state in the $\tfrac{N}{2}-1$ irrep (multiplicity labels ignored, as they are not disturbed by $\hat L_{j,q}$).\footnote{{From the standard definition of collective states, $\ket{\SCSangle}^{(J \neq J_\text{max})}$ is not well defined on its own due to multiplicities in lower irreps. However, the terms in the expressions above (Eq.~(\ref{eq:rotated_jump_term}) for example) are able to be treated as an outer product of vectors $\ket{\boldsymbol{\Omega}}$. Ref.~\cite{Forbes_Collective_States} showed that when restricting one's attention to collective properties of a system, collective states may be treated as true outer products.}}

We now need to evaluate $\hat L_{-1,-1} \ket{\Delta \SCSangle}$. To do so, we make use of 
    \begin{equation}
        \hat L_{-1,-1}=\sum_J\Lambda_{-1,-1}^{J,N}\hat T_{1,-1}^{J-1,J}, 
    \end{equation}
where $\Lambda_{-1,-1}^{J,N}$ is a coefficient given in Ref.~\cite{Forbes_Collective_States}, and $\hat T_{1,-1}^{N/2-1,N/2}$ is a spherical tensor operator that transforms under rotations as \cite{sakurai2014modern}, 
   \begin{equation} \label{eq:T_transform}
        \hat R(\SCSangle) \hat T^{\,J,J'}_{k,q}\, \hat R^\dagger(\SCSangle)
        =
        \sum_{q'=-k}^{+k}
        \mathscr{D}^{(k)}_{q' q}(\SCSangle)\,
        \hat T^{J,J'}_{k,q'} .
    \end{equation}
where $\mathscr{D}_{q,q'}^{(k)}(\SCSangle)$ is a Wigner-$D$ matrix. Combining these gives
    \begin{align} 
        &\hat L_{-1,-1} \ket{\Delta\SCSangle} \nonumber
        \\
        &= \hat R(\Delta\SCSangle) \big[\hat R^\dagger( \Delta\SCSangle) \hat L_{-1,-1}\hat R(\Delta \SCSangle)\big] \ket{0,0}
        \\
        &= \mathscr{D}_{-1,-1}^{(1)}(-\Delta\SCSangle) \hat R(\Delta\SCSangle)  \Lambda_{-1,-1}^{N/2,N} \hat T_{1,q}^{N/2-1,N/2}\ket{0,0}
        \\
        &= \mathscr{D}_{-1,-1}^{(1)}(-\Delta\SCSangle) \hat R(\Delta\SCSangle) \hat L_{-1,-1}\ket{0,0}
        \\
        &=\sqrt{N-1} \mathscr{D}_{-1,-1}^{(1)}(-\Delta\SCSangle) \hat R(\Delta\SCSangle) \ket{0,0}^{(N/2-1)}
        \\
        &= \sqrt{N-1} \cos^2 \Big(  \frac{\Delta \boldsymbol{\Omega}}{2} \Big) \ket{\Delta \SCSangle}^{(N/2-1)}
        \label{eq:LonSCS}
    \end{align}   
where $\ket{\Delta \SCSangle}^{(N/2-1)}$ is a SCS in the $\frac{N}{2}-1$ irrep, we used Eq.~\eqref{eq:Lon00}, and we substituted $\mathscr{D}_{-1,-1}^{(1)}(-\Delta\boldsymbol{\Omega}) = \cos^2 ( \frac{1}{2} \Delta \boldsymbol{\Omega})$. 

With Eq.~\eqref{eq:rotated_jump_term}, evaluated using Eq.~\eqref{eq:Lon00} and Eq.~\eqref{eq:LonSCS}, we find that the result of applying the jump map in Eq.~(\ref{eq:jumptermA}) to an arbitrary outer product of SCSs and projecting onto the $\frac{N}{2}-1$ irrep is
    \begin{align} \label{eq:final_state_symmetric_local_noise_app}
        \hat{P}_{N/2-1} & \bigg[ \sum_{j,q=-1}^1 \hat{L}_{j,q}  \ketbra{\SCSangle}{\SCSangle'} \hat{L}^\dagger_{j,q} \bigg] \hat{P}_{N/2-1} \nonumber 
        \\
        & = (N-1)
        \cos^2 \left( \frac{\Delta \boldsymbol{\Omega}}{2} \right)
        \overline{\ketbra{\SCSangle}{\SCSangle'}}^{(N/2-1)},
    \end{align}
with the $\overline{\ketbra{\SCSangle}{\SCSangle'}}^{(N/2-1)}$ lying in the space of collective states, see Eq.~\eqref{eq:collectivestateSCSop}.
When $\Delta \SCSangle$ is small, as is the case for bosonic code state imported via the HP approximation, $\cos^2 \Delta \boldsymbol{\Omega}/2 = 1 - \mathcal{O}( \Delta \boldsymbol{\Omega}^2 )$, and the distortion to ``coherences'' between the SCSs in superposition is small in the $\frac{N}{2}-1$ irrep.

\section{Preparing magic states for spin-GKP codes}
\label{sec:magic_states_spin_GKP}

A method for preparing magic states for the CV GKP code was introduced in Ref.~\cite{Ben_GKP_distillation} and later extended to qubit surface codes~\cite{gavriel2022transversalinjectionmethoddirect}, where it was dubbed \emph{transversal injection}. The key idea is to begin with an easy-to-prepare state (bosonic vacuum for GKP codes or an unencoded product state for the surface code) and use error correction to project it into the logical subspace to produce a state with distillable magic with high probability. These methods likewise can be used to produce magic states for HP spin codes; here, we adapt them specifically for the spin-GKP codes. In direct analogy with the CV GKP protocol, we propose a spin-GKP version that begins with the fully polarized state $\ket{\frac{N}{2},\frac{N}{2}} = \ket{0}^{\otimes N}$, which is a product state across the spins and also serves as the bosonic vacuum through the HP transformation.

While the analysis in Ref.~\cite{Ben_GKP_distillation} assumes ideal, infinite-energy GKP states, spin-GKP codes reside in a finite-dimensional Hilbert space, so the conclusions must be adapted accordingly.
In an ideal, infinite-energy setting, the projector onto the logical subspace is written as $\hat{\Pi} \propto \sum_m \hat{S}_m$, where $\{\hat{S}_m\}$ denotes the stabilizer group of the code. In the $N \rightarrow \infty$, the spin-GKP stabilizers coincide with those in \cref{eq:stabs_spinGKP}. For finite $N$, however, the ideal stabilizers are no longer exact. Following the approach of Royer \emph{et al.}~\cite{Shraddha_finite_energy} for finite-energy CV GKP states, we introduce damped spin-GKP stabilizers to account for this effect. Within the HP approximation, they are defined as
    \begin{equation}
        \hat{T}_p \to \hat{E} \hat{T}_p \hat{E}^{-1} \quad \text{and} \quad
        \hat{T}_q \to \hat{E} \hat{T}_q \hat{E}^{-1},
    \end{equation}
with the damping operator
    \begin{equation}
        \hat{E} \coloneqq \exp \big[-\Delta^2 \big( \tfrac{N}{2} - \hat{J}_z \big) \big].
    \end{equation}

\begin{figure}[b]
    \centering
    \includegraphics[width =0.9\columnwidth]{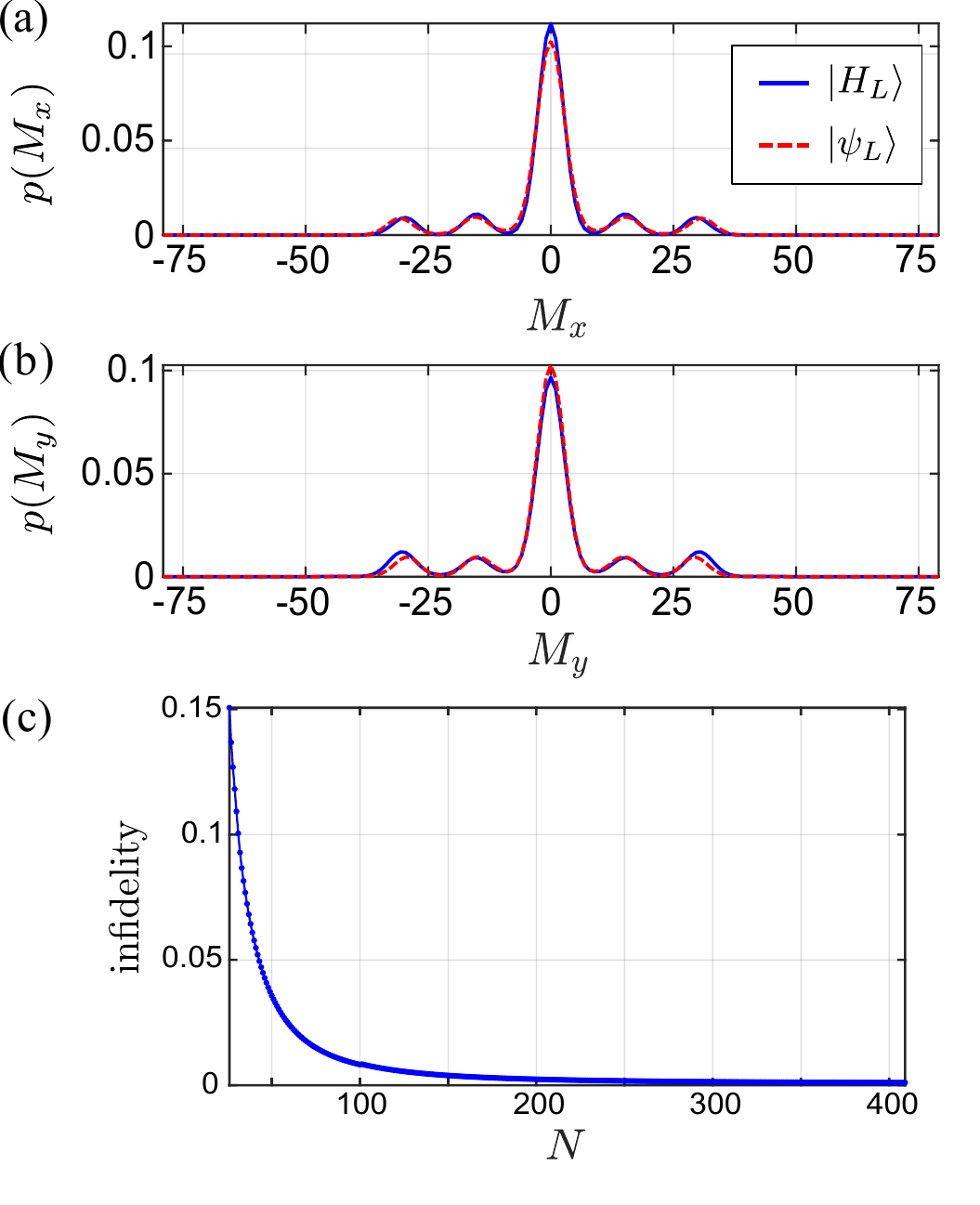}
    \caption{Performance of the resource-state preparation protocol for the spin-GKP code.
    (a) Transverse spin distributions for $N=160$. The solid blue line shows the intended magic state, and the dashed red shows the resource state in Eq.~\eqref{eq:resourcestate}.
    (c) Infidelity of the resource state as a function of $N$. }
    \label{fig:fig_magic_fidelity}
\end{figure}

Error correction with damped spin-GKP ancillae $\hat{E} \ket{\logic{0}}$ and $\hat{E} \ket{\logic{+}}$ produces the encoded resource state
    \begin{equation} \label{eq:resourcestate}
        \ket{ \logic \psi} \propto \hat{\Pi}_{q}\hat{\Pi}_{p}\ket{0}^{\otimes N},
    \end{equation}
where we have divided the damped codespace projector into two sectors as in Ref.~\cite{walshe2020},
    \begin{equation}
        \hat{\Pi}_q = \sum_m \hat{E} (\hat{T}_q)^m \hat{E}^{-1}, \quad
        \hat{\Pi}_p = \sum_m \hat{E} (\hat{T}_p)^m\hat{E}^{-1}.
    \end{equation}
Additionally, we post-select on the trivial syndrome. Our intention is simply to show that this procedure can succeed; a full analysis including success probabilities is beyond the scope of this work.

We compare the resource state $\ket{\logic \psi}$ with the target magic state $\ket{\logic H } = \cos (\tfrac{\pi}{8}) \ket{\logic 0} + \sin (\tfrac{\pi}{8}) \ket{\logic 1}$ in Fig.~\ref{fig:fig_magic_fidelity}.
Figures~\ref{fig:fig_magic_fidelity}(a) and (b) compare the transverse spin distributions for $N = 160$, which are nearly overlapping. A quantitative measure of performance is given by the infidelity $1 - | \braket{\logic \psi}{\logic{H}} |^2$, which we plot in Fig.~\ref{fig:fig_magic_fidelity}(c) as a function of $N$. As expected, the infidelity decreases monotonically with $N$, approaching the ideal magic state in the large $N$ limit. These results demonstrate that spin-GKP codes can natively host high-fidelity magic states, providing a practical route toward universal, fault-tolerant quantum computation.

\bibliography{reference}
\end{document}